\documentclass[english,aps,rmp,preprint,tightenlines,eqsecnum,longbibliography,notitlepage,showkeys]{revtex4-1}
\usepackage[T1]{fontenc}
\usepackage[latin9]{inputenc}
\usepackage{color}
\usepackage{babel}
\usepackage{amsmath}
\usepackage{amssymb}
\usepackage{stmaryrd}
\usepackage[unicode=true,
 bookmarks=true,bookmarksnumbered=false,bookmarksopen=false,
 breaklinks=true,pdfborder={0 0 0},pdfborderstyle={},backref=false,colorlinks=true]
 {hyperref}
\hypersetup{
 pdfauthor={François Konschelle},
 citecolor=blue, linkcolor=magenta,urlcolor=blue}

\makeatletter
\usepackage{eucal}

\DeclareMathOperator{\Tr}{Tr}

\DeclareMathOperator{\Pexp}{Pexp}

\makeatother

\begin{document}
\title{Gradient expansion of the non-Abelian gauge-covariant Moyal star-product}
\date{May 2015 - July 2017}
\author{François Konschelle}
\affiliation{Laboratoire de Mathématiques et Physique Théorique UMR 7350, Fédération
Denis Poisson FR 2964, Université François Rabelais de Tours, F-37200
Tours France}
\email{http://fraschelle.free.fr}

\keywords{Moyal star product ; non-Abelian gauge theory ; effective theory ;
kinetic theory ; quantum transport ; semi-classic and quasi-classic
approximation ; Wigner transform ; gradient expansion ; phase-space
quantum mechanics ; strict quantisation ; deformation quantisation
; topological field theory ;}
\begin{abstract}
Motivated by the recent developments of gauge-covariant methods in
the phase-space, a systematic method is presented aiming at the generalisation
of the Moyal star-product to a non-Abelian gauge covariant one at
any order. Such an expansion contains some dressing of the bare particle
model by the gauge-fields explicitly, and might serve as a drastically
simplifying tool for the elaborations of gauge-covariant quantum transport
models. In addition, it might be of fundamental importance for the
mathematical elaborations of gauge theory using the strict or deformation
quantisation principles. A few already known examples of quantum kinetic
theories are recovered without effort as an illustration of the power
of this tool. A gauge-covariant formulation taking into account possible
geometrical connections in both the position and momentum spaces is
also constructed at leading orders, with applications to the generation
of gauge-covariant effective theories in the phase-space. This paper
is devoted to the pedestrian elaboration of the gradient expansions.
Their numerous consequences will be explored in subsequent works.

\tableofcontents{}
\end{abstract}
\maketitle

\section{Introduction}

There have been considerable renewal and important developments in
the quasi-classic methods recently. They are all coming from many
temptatives to retain some gauge structures in the gradient expansion.
For instance, the prediction of a quantized transconductance in 2D
systems under magnetic field \citep{Thouless1982,Niu1985} -- the
quantum Hall effect phenomenology -- can be recovered in a quite
simple way using a gradient expansion \citep{Zubkov2016,Mera2017}.
Gradient expansion methods are also in use for the understanding of
topological phenomenologies in condensed-matter systems under interactions
\citep{Gurarie2011} and their bulk-boundary correspondance \citep{Essin2011}.
In superfluids, the gradient expansion naturally leads to a low-energy
effective action of topological nature as well \citep{Volovik1987,Volovik1988,Volovik1989}.

The quasi-classic methods take their roots in different attempts to
introduce quantum mechanics in the phase-space \citep{Wigner1932}
using the tools of statistical mechanics \citep{Moyal2008} or the
mathematical structure behind quantization of the position and momentum
coordinates \citep{Weyl1931,Groenewold1946}, reviewed in \citep{Hillery1984,Polkovnikov2009,Zachos2005,Curtright2014}.
Instead of using either the momentum $O\left(p_{1},p_{2}\right)=\left\langle p_{1}\right|\hat{O}\left|p_{2}\right\rangle $
or the position $O\left(x_{1},x_{2}\right)=\left\langle x_{1}\right|\hat{O}\left|x_{2}\right\rangle $
representations of a quantum operator $\hat{O}$ related via a Fourier
transform in both variables, Wigner introduced the so-called mixed-coordinate
Fourier transform, or Wigner transform, defined as
\begin{equation}
O\left(p,x\right)=\int d\mathfrak{x}\left[e^{-\mathbf{i}p\cdot\mathfrak{x}}O\left(x+\dfrac{\mathfrak{x}}{2},x-\dfrac{\mathfrak{x}}{2}\right)\right]=\int\dfrac{d\mathfrak{p}}{2\pi}\left[e^{\mathbf{i}\mathfrak{p}\cdot x}O\left(p+\dfrac{\mathfrak{p}}{2},p-\dfrac{\mathfrak{p}}{2}\right)\right]\label{eq:Wigner}
\end{equation}
in term of the center of mass $x=\left(x_{1}+x_{2}\right)/2$ and
quasi-momentum $p=\left(p_{1}+p_{2}\right)/2$. There is no change
of symbol between the Fourier representations $O\left(x_{1},x_{2}\right)$
or $O\left(p_{1},p_{2}\right)$ and the Wigner transformed $O\left(p,x\right)$.
When the correlation encoded in the relative coordinate $\mathfrak{x}=x_{1}-x_{2}$
or the momentum $\mathfrak{p}=p_{1}-p_{2}$ are supposed to be weak,
one may proceed to a perturbative expansion of the Wigner transform.
At zero-th order the expansion would give back the pure classical
results, and higher orders in the form of an expansion in powers of
derivatives (the so-called gradient expansion) then give the quantum
corrections to the statistical mechanics \citep{Wigner1932}.

It should not be surprising that the quasi-classic methods have profound
applications in statistical mechanics, and condensed matter problems
in particular. In fact, the later describes the phenomenology in the
real space of periodic materials having inherent band structures.
Also, many properties of a statistical system close to its equilibrium
state present generally long wavelength scalling, to be compared with
the atomistically ranged Fermi wavelength. Models taking into account
both the space and momentum variations are thus of great interest
there, and the gradient expansion is a powerfull tool to compute the
properties of a system close to its equilibrium. Methods of phase-space
quantum mechanics are in fact exploited in one way or another in many
problems of quantum transport \citep{Kadanoff1962,Abrikosov1963,Rammer1986,Langenberg1986,Rammer2008,Haug2010,b.kopnin}.
Moreover, the quasi-classic methods in the phase-space \citep{Agarwal1970e,Agarwal1970d,Agarwal1970c}
are also prominent in the understanding of the quantum-to-classical
transition and quantum optics \citep{Walls1994,Gardiner2004,Zachos2005}.

The success of the Wigner transform certainly lies in the complete
understanding of its algebra. Indeed, once the Wigner transforms of
two operators $O_{1,2}\left(x_{1},x_{2}\right)$ are known in the
form of $O_{1,2}\left(p,x\right)$, one can take the product of these
two operators in the phase-space following the general recipe of the
Moyal-Groenewold product
\begin{equation}
\int d\mathfrak{x}\int dy\left[e^{-\mathbf{i}p\cdot\mathfrak{x}}O_{1}\left(x+\dfrac{\mathfrak{x}}{2},y\right)O_{2}\left(y,x-\dfrac{\mathfrak{x}}{2}\right)\right]=O_{1}\left(p,x\right)\star_{0}O_{2}\left(p,x\right)\label{eq:Moyal-expansion}
\end{equation}
where the star $\star_{0}$-product (with $0$-index meaning there
is no gauge-potential associated to this product) has an explicit
form known to all orders \citep{Groenewold1946,Hillery1984} 
\begin{equation}
\star_{0}=\exp\left[\dfrac{\mathbf{i}}{2}\left(\partial_{x}^{\dagger}\partial_{p}-\partial_{p}^{\dagger}\partial_{x}\right)\right]
\end{equation}
with the notation convention that $O\partial_{x,p}^{\dagger}\equiv\partial_{x,p}O$,
i.e. the $\dagger$-derivative applies to the left. A gradient expansion
is thus the calculation of the different orders of the exponential
series of the $\star_{0}$ operation. This simplification tool allows
to quickly calculate many universal properties of effective theories
\citep{Volovik2003}, in addition to calculate kinetic models for
statistical systems \citep{Rammer2008,Haug2010}. For a recent overview
on the subject including many relevant references, see \citep{Curtright2014}.

There have been recent trends to try to generalise the gradient expansion
towards modern aspects of condensed matter problems. On the one hand,
a few attempts have been done towards a systematic expansion including
Berry phase effects of bands structure \citep{Shindou2005,Gosselin2006,Shindou2006,Shindou2008,Gosselin2008,Gosselin2007,Gosselin2008a,Gosselin2009,Wong2011,Wickles2013}
following the seminal works in \citep{Sundaram1999,Chang1995,Chang1996},
see reviews \citep{Xiao2010,Nagaosa2010}. Weyl semimetals also provide
an important platform to use such models \citep{Stephanov2012,Son2012,Stone2013,Son2013,Son2013a,Chernodub2017}.
On the other hand, spin-orbit effects can be handle as a non-Abelian
gauge theory \citep{Frohlich1993,Leurs2008,Berche2013}, which serves
as a guiding principle to the establishment of a kinetic theory when
charge and spin are treated in a gauge-covariant way \citep{Gorini2010,Bergeret2014,Konschelle2014}.
This later approach uses the so-called gauge-covariant Wigner transform,
defined as (to be justified further in section \ref{sec:Gauge-covariant-Moyal-star-product})
\begin{equation}
O\left(p,x\right)=\int d\mathfrak{x}\left[e^{-\mathbf{i}p\cdot\mathfrak{x}}U\left(x,x+\dfrac{\mathfrak{x}}{2}\right)O\left(x+\dfrac{\mathfrak{x}}{2},x-\dfrac{\mathfrak{x}}{2}\right)U\left(x-\dfrac{\mathfrak{x}}{2},x\right)\right]\label{eq:gauge-covariant-Wigner-transform}
\end{equation}
where the parallel transport operator of the gauge potential $A_{\alpha}$
(the greek indices run through the four space-time coordinates, and
the scalar product $p\cdot x=p^{\alpha}x_{\alpha}=\boldsymbol{p\cdot x}-\omega t$
can be eventually taken Lorentz invariant, with $\boldsymbol{p}$
the momentum vector in space, $\boldsymbol{x}$ the space position
vector, $\omega$ the frequency and $t$ the time coordinates) 
\begin{equation}
U\left(b,a\right)=\Pexp\left[\mathbf{i}\int_{a}^{b}dz\cdot A\left(z\right)\right]=\Pexp\left[\mathbf{i}\int_{0}^{1}ds\left[\left(b-a\right)\cdot A\left(\tau_{s}\right)\right]\right]\label{eq:parallel-transport}
\end{equation}
along the path-ordered straight line $\tau_{s}=a+\left(b-a\right)s$
from $a$ to $b$ collapses the possible gauge transformations of
the Wigner transform $O\left(p,x\right)$ to its position variable,
i.e. the function $O\left(p,x\right)$ transforms as $O^{\prime}\left(p,x\right)=R\left(x\right)O\left(p,x\right)R^{-1}\left(x\right)$
under the gauge transformation $A_{\alpha}^{\prime}\left(x\right)=R\left(x\right)A_{\alpha}\left(x\right)R^{-1}\left(x\right)-\mathbf{i}U\partial_{\alpha}U^{-1}$
of representation $R\left(x\right)$. Because \eqref{eq:gauge-covariant-Wigner-transform}
reduces to the usual Wigner transform \eqref{eq:Wigner} in the limit
of vanishing gauge fields $A\rightarrow0$, one did not change the
notation for the Wigner transform $O\left(p,x\right)$ of the operator
$O\left(x_{1},x_{2}\right)$. When not otherwise stated, a Wigner
transform $O\left(p,x\right)$ refers to the general definition \eqref{eq:gauge-covariant-Wigner-transform}.

When the gauge theory involves only Abelian groups, one can drastically
simplify the gauge-covariant Wigner transform \eqref{eq:gauge-covariant-Wigner-transform}
since in that case the two parallel transport operators \eqref{eq:parallel-transport}
collapses to a single one 
\begin{equation}
U_{\text{Abel.}}\left(x-\dfrac{\mathfrak{x}}{2},x+\dfrac{\mathfrak{x}}{2}\right)=\exp\left[-\mathbf{i}\int_{x-\mathfrak{x}/2}^{x+\mathfrak{x}/2}A\left(r\right)\cdot dr\right]
\end{equation}
such that the Wigner transform \eqref{eq:gauge-covariant-Wigner-transform}
reduces to the simpler expression 
\begin{equation}
O\left(p,x\right)=\int d\mathfrak{x}\left[e^{-\mathbf{i}p\cdot\mathfrak{x}}U_{\text{Abel.}}\left(x-\dfrac{\mathfrak{x}}{2},x+\dfrac{\mathfrak{x}}{2}\right)O\left(x+\dfrac{\mathfrak{x}}{2},x-\dfrac{\mathfrak{x}}{2}\right)\right]
\end{equation}
and some covariant formulations of the physical quantities in the
phase-space follow without much efforts, see \citep{Luttinger1951,Stratonovich1956,Kelly1964,Kubo1964,Langreth1966,Bialynicki-Birula1977,Serimaa1986,Javanainen1987,Bialynicki-Birula1991,Altshuler1992,Best1993,Kopnin1994,Levanda1994,Zachos1999,Levanda2001,Swiecicki2013}
for instance. A gauge-covariant Moyal expansion can even be established
in the Abelian case \citep{Muller1999a,Karasev2002,Karasev2003,Karasev2005,Mntoiu2004,Mantoiu2005,Iftimie2009},
with many applications in statistical physics problems \citep{Muller1999a,Lein2010}.

In the contrary, calculations using the non-Abelian version of the
gauge-covariant Wigner transform \eqref{eq:gauge-covariant-Wigner-transform}
are usually laborious and cumbersome. For some examples, one can refer
to \citep{Gorini2010,Bergeret2014,Konschelle2014} and the recent
pedagogical review \citep{Raimondi2016} for applications in condensed
matter, and \citep{Elze1986,Elze1986a,Winter1984,Winter1985} for
the original works developping a transport theory of the quark-gluon
plasma, or \citep{Elze1989,Weigert1991} for early reviews. A crucial
step to the understanding of its fundamental properties, and to the
simplification of its numerous modern uses, would be to establish
the non-Abelian gauge-covariant generalisation of the Moyal $\star$-product.
Unfortunately, it is already clear from the expression of the Abelian
gauge-covariant Moyal product, see e.g. \citep{Muller1999a}, that
the $\star$-product can not be writen in closed form if one wants
to preserve the gauge structure, unless defining extra mathematical
structures for which the exponential form of the Moyal product can
be obtained in a symbolic fashion, like in \citep{Bordemann1997,Bordemann1997a,Karasev2002,Karasev2003,Karasev2005,Mntoiu2004,Mantoiu2005,Iftimie2009}
where some involved mathematical structures are introduced. So we
would be happy to find a systematic method to generate the Moyal expansion
series order by order in a non-Abelian gauge-covariant way. Even if
this construction looks cumbersome at first, it will help simplifying
ulterior calculations, and understanding the gauge structure of the
models.

This study is dedicated to such a construction. I will present in
the following sections a method allowing to establish the first few
terms of the non-Abelian gauge-covariant Moyal $\star$-product associated
to the gauge-covariant Wigner transform
\begin{multline}
\int d\mathfrak{x}\int dy\left[e^{-\mathbf{i}p\cdot\mathfrak{x}}U\left(x,x+\dfrac{\mathfrak{x}}{2}\right)O_{1}\left(x+\dfrac{\mathfrak{x}}{2},y\right)O_{2}\left(y,x-\dfrac{\mathfrak{x}}{2}\right)U\left(x-\dfrac{\mathfrak{x}}{2},x\right)\right]=\\
=O_{1}\left(p,x\right)\star O_{2}\left(p,x\right)
\end{multline}
in the form
\begin{multline}
O_{1}\left(p,x\right)\star O_{2}\left(p,x\right)=O_{1}\left(p,x\right)O_{2}\left(p,x\right)+\dfrac{\mathbf{i}}{2}\left(\dfrac{\mathfrak{D}O_{1}}{\partial x^{\alpha}}\dfrac{\partial O_{2}}{\partial p_{\alpha}}-\dfrac{\partial O_{1}}{\partial p_{\alpha}}\dfrac{\mathfrak{D}O_{2}}{\partial x^{\alpha}}\right)+\\
-\dfrac{\mathbf{i}}{8}\left(F_{\alpha\beta}\left(x\right)\dfrac{\partial O_{1}}{\partial p_{\alpha}}\dfrac{\partial O_{2}}{\partial p_{\beta}}+2\dfrac{\partial O_{1}}{\partial p_{\alpha}}F_{\alpha\beta}\left(x\right)\dfrac{\partial O_{2}}{\partial p_{\beta}}+\dfrac{\partial O_{1}}{\partial p_{\alpha}}\dfrac{\partial O_{2}}{\partial p_{\beta}}F_{\alpha\beta}\left(x\right)\right)+\\
+\dfrac{1}{12}\left(\dfrac{\partial O_{1}}{\partial p_{\gamma}}\dfrac{\mathfrak{D}F_{\gamma\beta}}{\partial x^{\alpha}}\dfrac{\partial^{2}O_{2}}{\partial p_{\alpha}\partial p_{\beta}}+\dfrac{\partial^{2}O_{1}}{\partial p_{\alpha}\partial p_{\beta}}\dfrac{\mathfrak{D}F_{\gamma\beta}}{\partial x^{\alpha}}\dfrac{\partial O_{2}}{\partial p_{\gamma}}\right)+\\
+\dfrac{1}{24}\left[\dfrac{\mathfrak{D}F_{\gamma\beta}}{\partial x^{\alpha}},\dfrac{\partial^{2}O_{1}}{\partial p_{\alpha}\partial p_{\beta}}\dfrac{\partial O_{2}}{\partial p_{\gamma}}-\dfrac{\partial O_{1}}{\partial p_{\gamma}}\dfrac{\partial^{2}O_{2}}{\partial p_{\alpha}\partial p_{\beta}}\right]+\mathcal{O}\left(\partial^{4}\right)\label{eq:MOYAL-EXPANSION-GAUGE}
\end{multline}
up to the fourth order in gradient, and where we define the gauge
field 
\begin{equation}
F_{\alpha\beta}=\partial_{\alpha}A_{\beta}-\partial_{\alpha}A_{\beta}-\mathbf{i}\left[A_{\alpha},A_{\beta}\right]\label{eq:F-DEFINITION}
\end{equation}
 and the covariant derivative 
\begin{equation}
\dfrac{\mathfrak{D}O}{\partial x^{\alpha}}=\dfrac{\partial O}{\partial x^{\alpha}}-\mathbf{i}\left[A_{\alpha}\left(x\right),O\left(p,x\right)\right]\label{eq:D-DEFINITION}
\end{equation}
 applied to matrix. The fourth order term is also given in \eqref{eq:O-4}.

This result allows to drastically simplify the constructions of the
different versions of the gauge-covariant kinetic theories. Indeed,
one can compare the length of the demonstration of \eqref{eq:MOYAL-EXPANSION-GAUGE}
given below with the relative simplicity of its utilisation in section
\ref{sec:Moyal-product-in-action}. It might also be of importance
in the understanding of the so-called deformation quantisation \citep{Bayen1977,Bayen1978a,Bayen1978,Kontsevich1994}
or strict quantisation \citep{Rieffel1993,Rieffel1994,Landsman1998,Landsman2005},
since the gauge-covariant Moyal expansion \eqref{eq:MOYAL-EXPANSION-GAUGE}
will be shown to dress the bare theory with a gauge structure in section
\ref{subsec:Non-commutative-dressing}. Indeed, one way of applying
the gauge-covariant Moyal product is to start from the action of a
bare particle, for instance the Dyson's propagator 
\begin{equation}
G^{-1}\left(x_{1},x_{2}\right)=\left(\mathbf{i}\partial_{t}+A_{0}+\dfrac{\left(\partial_{x}-\mathbf{i}A\right)^{2}}{2m}+\mu\right)\delta\left(x_{1}-x_{2}\right)
\end{equation}
for a Galilean relativistic free particle in a gas of chemical potential
$\mu$, and to get the gauge-covariant Wigner transform of it, which
is nothing but 
\begin{equation}
G^{-1}\left(p,x\right)=\omega-\dfrac{p^{2}}{2m}+\mu
\end{equation}
i.e. the bare classical action. Then the gauge-covariant Moyal expansion
\eqref{eq:MOYAL-EXPANSION-GAUGE} applied to the Dyson's equation
restaures the known transport theories in normal metals \citep{Gorini2010}
and superconductors \citep{Konschelle2014,Bergeret2014} dressed with
Abelian and/or non-Abelian gauge fields. This will be the subject
of section \ref{subsec:Quantum-kinetic-theory}, where a few already
known gauge-covariant transport equations will be derive using \eqref{eq:MOYAL-EXPANSION-GAUGE}
directly instead of dealing with the gauge-covariant Wigner transform
\eqref{eq:gauge-covariant-Wigner-transform} in a cumbersome way.

The gradient expansion \eqref{eq:MOYAL-EXPANSION-GAUGE} also leads
to direct applications in deriving universal properties of statstical
systems. In section \ref{subsec:Topology-in-the-momentum-space} we
derive the Chern-number quantisation of the transconductance in a
non-interacting system, and its generalisation to interacting system
in the form of a topological invariant for the Green's functions.
This well known result \citep{Thouless1982,Niu1985} has been recovered
recently using a gauge-flat gradient expansion \citep{Zubkov2016,Mera2017}
which, by properties of the Abelian electromagnetic gauge field, can
be restated in a gauge-covaraint fashion quite easilly. We will show
that the gradient expansion \eqref{eq:MOYAL-EXPANSION-GAUGE} directly
gives the result without effort. The appearance of Pontryagin number
in superfluids, first established in a series of paper by Volovik
and Yakovenko \citep{Volovik1987,Volovik1988,Volovik1989}, is also
reviewed in section \ref{subsec:Topology-in-the-momentum-space} in
a quite straightforward way.

We then turn to the explicit demonstration of the gauge-covariant
gradient expansion \eqref{eq:MOYAL-EXPANSION-GAUGE} up to the fourth
order in section \ref{sec:Gauge-covariant-Moyal-star-product}. The
demonstration uses a trick introduced in \citep{Weigert1991} of translating
the gauge-covariant Wigner transform towards the Fock-Schwinger gauge
(also called relativistic Poincaré, radial, quasi-canonical or fixed-point
gauge fixing). In this gauge, the product of two operators can be
recast in the form of the flat-space Moyal expansion \eqref{eq:Moyal-expansion}.
Then one has to transport back this equation to the gauge-covariant
formulation of phase-space functions. That is, one has to understand
how the different orders of space derivatives in the Fock-Schwinger
gauge translate to the gauge-covariant Wigner transform. This is the
work accomplished in section \ref{sec:Gauge-covariant-Moyal-star-product}.

Since there have been proposals to get some gauge-covariant expressions
in the full phase-space when both $D_{x}=\partial_{x}-\mathbf{i}A_{x}\left(x\right)$
and $D_{p}=\partial_{p}-\mathbf{i}A_{p}\left(p\right)$ covariant
derivatives pop out of the non-covariant phase-space method of effective
diagonalisation of the band structure \citep{Shindou2005,Gosselin2006,Shindou2006,Shindou2008,Gosselin2008,Gosselin2007,Gosselin2008a,Gosselin2009,Wong2011,Wickles2013},
a few words about the possible realisations of the gradient expansion
retaining the full geometry of the phase-space might be welcome. In
fact, such an approach nicely generalises the demonstration done in
section \ref{sec:Gauge-covariant-Moyal-star-product}. Section \ref{sec:Gauge-covariant-gradient-expansion-phase-space}
is dedicated to such a construction, generalising and unifying a few
previous works.

\subsection{Conventions\label{subsec:Conventions}}

Before turning to the application of the gauge-covariant Moyal expansion
\eqref{eq:MOYAL-EXPANSION-GAUGE}, a few words about mathematical
rigor are in order. Indeed, there will be no attempt toward rigorously
define the conditions under which the expansion converges, nor to
the rigorous definition of the gauge-covariant Wigner transform and
its associated Weyl mapping from the classical to the quantum world.
Reader interested in more rigorous construction might consult e.g.
\citep{Estrada1989,Landsman1993a,Robson1994,Alvarez-Gaume2001a,Szabo2003,Wallet2009}
for general construction, and especially \citep{Bordemann1997,Bordemann1997a}
for the construction of a covariant star-product, and references therein.
\citep{Sugimoto2007,Sugimoto2012} also claim to have used a non-Abelian
gauge-covariant method to extract the conservation laws of the spin
degree of freedom, but I found nothing to compare with the approach
presented below. In contrary, a covariant method have been proposed
to compute the effective action in the high-energy sector \citep{Pletnev1999,Pletnev2001,Salcedo2007}
following similar recipes than I. Nevertheless, these studies detail
only the situation when the gauge field has effect in the position
space, and there is no explicit construction of the star product as
I do below.In addition, a simpler examination of the method might
be welcome. 

In the following, I propose a pedestrian approach to the establishment
of the first few terms of the Moyal expansion up to the fourth order,
and the results of this study can be seen as a quasi-empirical construction,
hopefully usefull for tackling modern problems of condensed matter,
more generally for any problem when quantisation of the gauge-fields
is not required. The aim is to provide a demonstration of \eqref{eq:MOYAL-EXPANSION-GAUGE}
that a graduate student having basic knowledge on gauge theory and
Moyal product \eqref{eq:Moyal-expansion} can follow. 

A few words on the notation convention: The space-momentum Wigner
correspondance is done using the definition 
\begin{equation}
O\left(\boldsymbol{p},\boldsymbol{x}\right)=\int d\boldsymbol{\mathfrak{x}}\left[e^{-\mathbf{i}\boldsymbol{p\cdot\mathfrak{x}}}O\left(\boldsymbol{x}+\dfrac{\boldsymbol{\mathfrak{x}}}{2},\boldsymbol{x}-\dfrac{\boldsymbol{\mathfrak{x}}}{2}\right)\right]=\int d\mathfrak{p}\left[e^{\mathbf{i}\boldsymbol{x\cdot\mathfrak{p}}}O\left(\boldsymbol{p}+\dfrac{\mathfrak{p}}{2},\boldsymbol{p}-\dfrac{\mathfrak{p}}{2}\right)\right]
\end{equation}
and the time-frequency correspondance using the Wigner transform follows
from the convention
\begin{equation}
O\left(\omega,t\right)=\int d\mathfrak{t}\left[e^{\mathbf{i}\omega\mathfrak{t}}O\left(t+\dfrac{\mathfrak{t}}{2},t-\dfrac{\mathfrak{t}}{2}\right)\right]=\int d\mathfrak{w}\left[e^{-\mathbf{i}\mathfrak{w}t}O\left(\omega+\dfrac{\mathfrak{w}}{2},\omega-\dfrac{\mathfrak{w}}{2}\right)\right]
\end{equation}
where the integrated variables, corresponding to relative coordinates,
are noted with German alphabet. The covariant derivatives are writen
as 
\begin{equation}
\dfrac{Df}{\partial x^{\alpha}}=\dfrac{\partial f}{\partial x^{\alpha}}-\mathbf{i}A_{\alpha}\left(x\right)f\;\Rightarrow\begin{cases}
\dfrac{Df}{\partial t} & =\dfrac{\partial f}{\partial t}-\mathbf{i}A_{0}\left(x\right)f\\
\\
\dfrac{Df}{\partial\boldsymbol{x}} & =\dfrac{\partial f}{\partial\boldsymbol{x}}-\mathbf{i}\boldsymbol{A}\left(x\right)f
\end{cases}
\end{equation}
for a function $f\left(x,t\right)$ and space-time dependent gauge
potential $A_{\alpha}\left(x,t\right)$. The correspondence is the
same for 
\begin{equation}
\dfrac{\mathfrak{D}F}{\partial x^{\alpha}}=\dfrac{\partial F}{\partial x^{\alpha}}-\mathbf{i}\left[A_{\alpha}\left(x\right),F\left(p,x\right)\right]
\end{equation}
for a matrix (this covariant derivative appears only in the phase
space). As a consequence, the Moyal $\star_{0}$ product reads
\begin{equation}
\star_{0}=\exp\left[\dfrac{\mathbf{i}}{2}\left(\dfrac{\partial^{\dagger}}{\partial\boldsymbol{x}}\dfrac{\partial}{\partial\boldsymbol{p}}-\dfrac{\partial^{\dagger}}{\partial\boldsymbol{p}}\dfrac{\partial}{\partial\boldsymbol{x}}-\dfrac{\partial^{\dagger}}{\partial t}\dfrac{\partial}{\partial\omega}+\dfrac{\partial^{\dagger}}{\partial\omega}\dfrac{\partial}{\partial t}\right)\right]\label{eq:Moyal-relativistic}
\end{equation}
where we restate the dimension by multiplying every $\boldsymbol{p}$-derivative
with a $\hbar$. The context and the notations make clear whether
we use the complete relativistic Wigner representation $O\left(p,x,\omega,t\right)$,
the space-momentum one $O\left(p,x\right)$ (also writen $O\left(p,x,t_{1},t_{2}\right)$
or $O\left(p,x,\omega_{1},\omega_{2}\right)$ for time-dependent problems)
or the time-frequency one $O\left(\omega,t\right)$ (also writen $O\left(\omega,t,x_{1},x_{2}\right)$
or $O\left(\omega,t,p_{1},p_{2}\right)$ for space-dependent problems). 

I put $\hbar=1$ everywhere. To restate the dimension, one can add
a $\hbar$ term in front of all $\partial_{p}$-derivative. The gauge
structure has natural dimension of inverse of length.

The notations adapt themselves to the situation of a $p$-dependent
gauge potential $A^{\alpha}\left(p\right)$ and covariant derivative
$\mathfrak{D}^{\alpha}F=\partial^{\alpha}F-\mathbf{i}\left[A^{\alpha}\left(p\right),F\left(p,x\right)\right]$,
with the convention $\partial^{\alpha}f=\partial f/\partial p_{\alpha}$
for the momentum derivative. See section \ref{subsec:Momentum-like-gauge-covariant-gradient-expansion}
for more details. 

There is no try to discuss the symplectic structure of the phase-space,
but positions of the indices are obvious in section \ref{subsec:Phase-space-gradient-expansion}.

\subsection{Emergence of Gauge structures\label{subsec:Emergence-of-Gauge-structures}}

Before entering the heart of the calculations, I would like to introduce
some gauge structures naturally emerging in condensed matter systems.
The reason is first of all pedagogical. I will discuss essentially
classical systems in the following, and the star product \eqref{eq:MOYAL-EXPANSION-GAUGE}
will be shown to dress the classical theory with some gauge structures.
That is, the covariant star product exhibits some gauge connections
and gauge fields, but their origin and exact expression in terms of
microscopic degrees of freedom are hidden in the writting of $A_{\alpha}$
(in the derivatives $\mathfrak{D}_{\alpha}$) and $F_{\alpha\beta}$.
It might then appear to the reader that this dressing comes from pure
black magic, or even worse, that the proposed construction is of pure
scholar interest without any physical ground. To explain why this
is not the case is the purpose of this section, aiming at introducing
the origin of the gauge structure from microscopic considerations.

Indeed, the characteristics of the gauge structures of interest for
semi-classic arguments (e.g. which group is relevant for the description
of a given phenomenology) inherited from \eqref{eq:MOYAL-EXPANSION-GAUGE}
are in practise emerging from microscopic considerations. In a way,
the semi-classic expansion I propose below inherits the geometry of
the quantum state, but the method I will develop is blind with respect
to this microscopic origin once the quantities $A_{\alpha}$, $\mathfrak{D}_{\alpha}$
and $F_{\alpha\beta}$ are known. As a matter of fact, physical results
can be expressed in terms of the gauge connections and gauge fields
only, without any mention of their origin, so to write everything
in terms of $A_{\alpha}$, $\mathfrak{D}_{\alpha}$ and $F_{\alpha\beta}$
should not be such surprising. In this section, I introduce quickly
the associated microscopic considerations, and refer to the relevant
studies for more details. Then in the sequels of this paper, I will
suppose that one already knows the specificity of the gauge structure,
i.e. I will suppose one knows $A_{\alpha}$, $\mathfrak{D}_{\alpha}$
and $F_{\alpha\beta}$ from the outset. I will then show that a covariant
gradient expansion exists, and built it in terms of the gauge connection
and gauge fields without any more mention of their microscopic origin.

As a first example, nothing better than the celebrated electromagnetism.
In that situation, it comes as a natural reflex to dress the bare
particle via the minimal substitution $p_{i}\rightarrow p_{i}-\mathcal{A}_{i}^{\text{e.m.}}$,
with $\mathcal{A}_{i}^{\text{e.m.}}=eA_{i}$ the so-called vector
gauge potential, with gauge charge $e$. In that case, the associated
gauge field is Abelian and is usually called the magnetic field in
the space sector $F^{ij}=\varepsilon^{ijk}B^{k}$, when latin indices
run along the three space dimensions, $i\in\left\{ 1,2,3\right\} $
and $\varepsilon^{ijk}$ is the completely antisymmetric symbol, and
the electric field $F^{0i}=E^{i}$ in the time sector having $0$
for index. When putting $\mathcal{A}_{\alpha}^{\text{e.m.}}$ instead
of $A_{\alpha}$ in \eqref{eq:MOYAL-EXPANSION-GAUGE}-\eqref{eq:D-DEFINITION},
the Moyal product will dress the bare classical system with the electromagnetic
interaction. Many simplifications follows from the Abelian nature
of electromagnetism.

More involved gauge structures, namely non-Abelian ones, naturally
show up in the realm of high energy physics. A pedagogical introduction
to such physics can be find in \citep{Aitchison2004}. Once more time,
substitution of the gauge structure in \eqref{eq:MOYAL-EXPANSION-GAUGE}-\eqref{eq:D-DEFINITION}
will dress the bare theory with the adapted geometry. I now come back
to condensed matter problems, and introduce (perhaps) less trivial
gauge structures in the real and momentum space.

What appeared recently as an interesting change of paradigm in condensed
matter problems was the possibility to treat complicated spin couplings
using the language of gauge theory \citep{Frohlich1993,Leurs2008,Berche2013}.
For instance, the spin-orbit interaction naturally appears in a free
electron gas of effective mass $m$, momentum $p$ and chemical potential
$\mu$ without inversion symmetry following the Hamiltonian description
\begin{equation}
H_{\text{s.o.}}=\dfrac{p^{2}}{2m}-\mu-\alpha_{i}^{a}\left(x\right)\dfrac{\sigma^{a}p_{i}}{2}=\dfrac{\left(p_{i}-m\alpha_{i}^{a}\left(x\right)\sigma^{a}/2\right)^{2}}{2m}-\mu_{0}\label{eq:model-spin-orbit}
\end{equation}
where repeated indices are summed, and $\sigma^{a}$ are the Pauli
matrices which span the spin algebra. The spin-orbit interaction is
encoded in the $p$-independent tensor $\alpha_{i}^{a}$. In the second
expression $\mu_{0}=\mu-m\left(\alpha_{i}^{a}\right)^{2}/4$ is a
rescalling of the chemical potential which has no physical consequence.
Now, we can formally define $\mathcal{A}_{i}^{\text{s.o.}}=m\alpha_{i}^{a}\sigma^{a}/2$
as a gauge potential. The associated gauge structure corresponds to
the geometry of rotations, here $\text{SU}\left(2\right)$ for electronic
spin, hence $\mathcal{A}_{i}^{\text{s.o.}}$ is non-Abelian. When
$A_{i}$ is replaced with $\mathcal{A}_{i}^{\text{s.o.}}=m\alpha_{i}^{\alpha}\sigma^{a}/2$
in the Moyal expansion \eqref{eq:MOYAL-EXPANSION-GAUGE}, the gradient
expansion will describe the phenomenology of the spin interaction
in condensed matter systems. Note that the usual shift $H\rightarrow H-\mathcal{A}_{0}^{\text{Z.}}$
with $\mathcal{A}_{0}^{\text{Z.}}=B^{a}\sigma^{a}/2$ can describe
the usual Zeeman coupling between the spin and the electromagnetic
field. Such a situation can also be described following the present
study: it is sufficient to either deal with quasi-static situations
when $\mathcal{A}_{0}^{\text{Z.}}$ is time-independent, or to extend
the gradient expansion to deal with adiabatic contributions, when
one uses both the position-momentum and the time-frequency Wigner
transformations, for instance \eqref{eq:Moyal-relativistic} when
there is no gauge-field. The notations introduced in section \ref{subsec:Conventions}
explain how to deal with these situations.

The main merit of the gauge theory is then to be abble to deal, e.g.,
with charge and spin degrees of freedom using the substitution $A_{i}=\mathcal{A}_{i}^{\text{e.m.}}+\mathcal{A}_{i}^{\text{s.o.}}$
for instance. This is the strategy followed in order to deal with
spintronics applications either in semi-conducting systems \citep{Gorini2010,Raimondi2016}
or superconducting systems \citep{Bergeret2014,Konschelle2014}.

As the last example of this short review, I would like to introduce
the gauge structure in the momentum space, which attracted much interest
in the later years, especially in the field of topological materials
\citep{Sundaram1999,Berard2004,Stephanov2012,Son2012}, see also \citep{Xiao2010,Nagaosa2010,Charge2016}
for reviews. To understand how a gauge structure emerges from the
momentum space, let us discuss a simplified model, with Hamiltonian
\begin{equation}
H=H_{0}\left(p\right)+V_{0}x\label{eq:model-band-perturbation}
\end{equation}
where $H_{0}\left(p\right)$ can be thought as the Hamiltonian representation
of a band structure, and $V_{0}$ some perturbative potential. The
position dependency of the perturbation is supposed to be linear,
so $V_{0}$ is a constant matrix in \eqref{eq:model-band-perturbation}.
The model \eqref{eq:model-band-perturbation} is related to physical
systems of interest: the perturbation $V_{0}x$ mimics a constant
electrostatic contribution applied to a solid state or an atom, for
instance, or some strain applied to the atom lattice.

Suppose next that the unitary transformation $R\left(p\right)$ diagonalises
the band structure, i.e. $R\left(p\right)H_{0}\left(p\right)R^{\dagger}\left(p\right)=\mathcal{E}\left(p\right)$
is a diagonal matrix. Then one has 
\begin{equation}
UHU^{\dagger}=\mathcal{E}\left(p\right)+\tilde{V}_{0}\left(p\right)R\left(p\right)xR^{\dagger}\left(p\right)
\end{equation}
where the perturbative matrix $\tilde{V}_{0}\left(p\right)=R\left(p\right)V_{0}R^{\dagger}\left(p\right)$
might have acquired a momentum dependency under the transformation
(since we are in the interaction representation now). Since $x$ and
$p$ do not commute in quantum mechanics, one has (a similar argument
has been used in \citep{Gosselin2006}) 
\begin{equation}
R\left(p\right)xR^{\dagger}\left(p\right)=x+R\left(p\right)\left[x,R^{\dagger}\left(p\right)\right]=x+\mathbf{i}R\dfrac{\partial R^{\dagger}}{\partial p}\label{eq:model-band-phase-shift}
\end{equation}
using the formal relation $\left[x,f\left(p\right)\right]=\mathbf{i}$$\partial_{p}f$
for any analytic function $f$ of the momentum, and with $\left[x,p\right]=\mathbf{i}$.
Then under a change of representation of the model \eqref{eq:model-band-perturbation},
the system acquires a shift $\tilde{V}_{0}R\partial_{p}R^{\dagger}$
with origin in the momentum space. This is not yet a genuine gauge
structure, since the shift $R\partial_{p}R^{\dagger}$ is curvature-free
and has no physical effect (in the language of gauge theory, such
a contribution is called a pure gauge). Nevertheless, if one restricts
the total Hilbert space span by the model \eqref{eq:model-band-perturbation}
in a way or an other using the projection $\mathcal{P}$, usually
in the low energy sector of an effective theory, the projected contribution
$\mathcal{A}^{\text{p}}=\mathcal{P}\left[R\partial_{p}R^{\dagger}\right]$
to this restriction might become a true connection, and might have
influences on the dynamics of the system. This is how the gauge structure
emerges in the momentum space. One can deal also with the time-dependent
Schrödinger equation and the adiabatic theorem (instead of the effective
theory in the band structure, see \citep{Moore2017} for a recent
short review), in which case a frequency-like gauge potential appears,
in addition to the momentum-like gauge structure. Once the gauge-potentials
$\mathcal{A}^{\text{p}}=\mathcal{P}\left[R\partial_{p}R^{\dagger}\right]$
is inserted into the covariant gradient expansion of section \ref{subsec:Momentum-like-gauge-covariant-gradient-expansion}
(where we will deal with momentum covariant structures), one no more
need the reference to the microscopic model in \eqref{eq:model-band-phase-shift},
that is, there is no more need to deal with non-commuting phase-space
variables as these later ones will be naturally encapsulated in the
Moyal expansion.

Note in passing that the shift \eqref{eq:model-band-phase-shift}
might not exist for interactions of polynomial higher orders in position,
since the commutator (see \citep{Transtrum2005} for the proof) 
\begin{equation}
\left[f\left(p,x\right),g\left(p,x\right)\right]=\sum_{k=1}^{\infty}\dfrac{\left(-\mathbf{i}\right)^{k}}{k!}\left(\dfrac{\partial^{k}g}{\partial x^{k}}\dfrac{\partial^{k}f}{\partial p^{k}}-\dfrac{\partial^{k}f}{\partial x^{k}}\dfrac{\partial^{k}g}{\partial p^{k}}\right)\label{eq:model-commutator}
\end{equation}
of two analytic functions of the momentum and the position gets a
clear gauge structure (once projected to some effective sub-spaces)
only at linear order in space and/or momentum. Nevertheless, the apparent
similarity between \eqref{eq:model-commutator} and the Moyal commutator
$\left[f,g\right]_{\star_{0}}=f\star_{0}g-g\star_{0}f$ using the
$\star_{0}$-product \eqref{eq:Moyal-expansion} tends to show how
phase-space gauge structures might emerge at the leading order of
a semi-classic expansion of effective models, as discussed in \citep{Shindou2005,Gosselin2006,Shindou2006,Shindou2008,Gosselin2008,Gosselin2007,Gosselin2008a,Gosselin2009,Wong2011,Wickles2013}
in different contexts, usually not using covariant Wigner transform
in condensed matter contexts.

We have thus seen that linear-in-momentum interaction naturally leads
to an emergent position-dependent gauge structure (the example \eqref{eq:model-spin-orbit}
of spin-orbit coupling), whereas a linear-in-posititon interaction
naturally leads to an emergent momentum-dependent gauge structure
in effective models (the example \eqref{eq:model-band-phase-shift}
is a prototype of such models). In the following, we will neglect
the origin of the gauge structure. Instead, we will suppose there
is a gauge structure at the microscopic level, and we will learn how
to include this gauge structure at the semi-classic level in terms
of $A_{\alpha}$ and $F_{\alpha\beta}$ without worrying anymore about
their origin. Section \ref{subsec:Phase-space-gradient-expansion}
introduces a covariant gradient expansion valid in the entire phase-space
when both a position dependent and a momentum dependent gauge potentials
are present.

\section{The covariant Moyal product in action\label{sec:Moyal-product-in-action}}

In this section, a few examples of the uses of the Moyal expansion
\eqref{eq:MOYAL-EXPANSION-GAUGE} are presented, in order to show
how powerfull this tool is. Theses examples are all already known
in the literature, but the treatment is greatly simplified by the
knowledge of the $\star$-product \eqref{eq:MOYAL-EXPANSION-GAUGE}.
In the following, we mainly review the situation in condensed matter
problems, where the gauge-covariant Wigner transform proved to be
a convenient tool to unify the treatment of charge and spin degrees
of freedom in terms of gauge principles applied to quantum kinetic
equations \citep{Gorini2010,Konschelle2014,Bergeret2014,Raimondi2016},
see section \ref{subsec:Quantum-kinetic-theory}. In addition, we
easilly generalises the phenomenology of the quantum Hall effect to
non-Abelian structure (namely for spin current) in section \ref{subsec:Topology-in-the-momentum-space}.
We start by explaining one important property of the Moyal product
\eqref{eq:MOYAL-EXPANSION-GAUGE}: it dresses the classical theory
with quantum corrections coming from the gauge structure in section
\ref{subsec:Non-commutative-dressing}.

\subsection{Non-commutative dressing by the gauge structure\label{subsec:Non-commutative-dressing}}

Let us start with the expansion of the momentum operator, namely,
the calculation of the star product \eqref{eq:MOYAL-EXPANSION-GAUGE}
with $O_{1}\left(p,x\right)=p_{\alpha}$: 
\begin{equation}
p_{\alpha}\star O_{2}\left(p,x\right)=p_{\alpha}O_{2}\left(p,x\right)-\dfrac{\mathbf{i}}{2}\dfrac{\mathfrak{D}O_{2}}{\partial x^{\alpha}}-\dfrac{\mathbf{i}}{8}\left(3F_{\alpha\beta}\dfrac{\partial O_{2}}{\partial p_{\beta}}+\dfrac{\partial O_{2}}{\partial p_{\beta}}F_{\alpha\beta}\right)+\cdots\label{eq:p-star-G}
\end{equation}
up to the first order for convenience. We realise that this is nothing
but the gauge-covariant quasi-classic expansion of the covariant derivative
obtained after long algebra in e.g. \citep{Elze1989,Konschelle2014}
(up to differences in conventions). So we have just shown that, 
\begin{multline}
\int d\mathfrak{r}\left[e^{-\mathbf{i}p\cdot\mathfrak{x}}U\left(x,x+\dfrac{\mathfrak{r}}{2}\right)D_{\alpha}\left(x+\dfrac{\mathfrak{x}}{2}\right)O\left(x+\dfrac{\mathfrak{x}}{2},x-\dfrac{\mathfrak{x}}{2}\right)U\left(x-\dfrac{\mathfrak{x}}{2},x\right)\right]=\\
=p_{\alpha}\star O\left(p,x\right)
\end{multline}
namely, the gauge-covariant gradient expansion of the covariant derivative
$D_{\alpha}\left(x\right)O\left(x,y\right)=\partial O/\partial x^{\alpha}-\mathbf{i}A_{\alpha}\left(x\right)O\left(x,y\right)$
equals the $\star$-product between $p_{\alpha}$ and the Wigner transform
$O\left(p,x\right)$ of $O\left(x_{1},x_{2}\right)$. In particular,
one notes that, formaly 
\begin{equation}
\int d\mathfrak{r}\left[e^{-\mathbf{i}p\cdot\mathfrak{x}}U\left(x,x+\dfrac{\mathfrak{r}}{2}\right)D_{\alpha}\left(x+\dfrac{\mathfrak{x}}{2}\right)U\left(x-\dfrac{\mathfrak{x}}{2},x\right)\right]=p_{\alpha}
\end{equation}
in the Moyal product \eqref{eq:MOYAL-EXPANSION-GAUGE}. Hence the
$\star$-product dresses the theory with covariant contributions $\mathfrak{D}_{\alpha}$
and $F_{\alpha\beta}$ coming from the gauge structure of the microscopic
theory, even if one starts with a bare situation, i.e. the $\star$
operation in \eqref{eq:p-star-G} takes the bare classical $p_{\alpha}$
and $O\left(p,x\right)$ as inputs and outputs a covariant expression.

This is perhaps the most important property of \eqref{eq:MOYAL-EXPANSION-GAUGE}:
it maps two gauge-less classical quantities to their semi-classic
counterparts and provides the dressing by the internal gauge structure
of the microscopic theory.

In the same way, one has 
\begin{multline}
\int d\mathfrak{r}\left[e^{-\mathbf{i}p\cdot\mathfrak{x}}U\left(x,x+\dfrac{\mathfrak{r}}{2}\right)O_{1}\left(x+\dfrac{\mathfrak{x}}{2},x-\dfrac{\mathfrak{x}}{2}\right)D_{\alpha}^{\dagger}\left(x-\dfrac{\mathfrak{x}}{2}\right)U\left(x-\dfrac{\mathfrak{x}}{2},x\right)\right]=\\
=O_{1}\left(p,x\right)\star p_{\alpha}=p_{\alpha}O_{1}\left(p,x\right)+\dfrac{\mathbf{i}}{2}\dfrac{\mathfrak{D}O_{1}}{\partial x^{\alpha}}+\dfrac{\mathbf{i}}{8}\left(F_{\alpha\beta}\dfrac{\partial O_{1}}{\partial p_{\beta}}+3\dfrac{\partial O_{1}}{\partial p_{\beta}}F_{\alpha\beta}\right)+\cdots\label{eq:G-star-p}
\end{multline}
with $D_{\alpha}^{\dagger}\left(x\right)=\partial^{\dagger}/\partial x^{\alpha}+\mathbf{i}A_{\alpha}\left(x\right)$.
Note the change in sign in front of the gauge-field term due to its
antisymmetry, and the balance between the prefactors of $\partial_{p}O_{1}$
and $\partial_{p}O_{2}$ in expressions \eqref{eq:p-star-G} and \eqref{eq:G-star-p}.

To see the emergence of the non-commutative geometry associated with
\eqref{eq:MOYAL-EXPANSION-GAUGE}, we can also calculate the following
Moyal commutation relations 
\begin{align}
\left[p_{\alpha},p_{\beta}\right]_{\star} & =p_{\alpha}\star p_{\beta}-p_{\beta}\star p_{\alpha}=-\mathbf{i}F_{\alpha\beta}\left(x\right)\nonumber \\
\left[x^{\alpha},p_{\beta}\right]_{\star} & =x^{\alpha}\star p_{\beta}-p_{\beta}\star x^{\alpha}=\mathbf{i}\delta_{\beta}^{\alpha}\nonumber \\
\left[x^{\alpha},x^{\beta}\right]_{\star} & =x^{\alpha}\star x^{\beta}-x^{\beta}\star x^{\alpha}=0\label{eq:Moyal-commutation-non-commutative}
\end{align}
where the gauge field appears in front of the momentum derivative.
One more time, such a commutation relation is usual in the quantum
world since $\left[D_{\alpha}\left(x\right),D_{\beta}\left(x\right)\right]=-\mathbf{i}F_{\alpha\beta}$,
inducing a non-commutative geometry in the phase space. Such possibilities
have been well explored in the case of the quantum Hall effect, see
e.g. \citep{Bellissard1994} and references therein. Here the gauge
field can be non-Abelian as well. Of course \eqref{eq:MOYAL-EXPANSION-GAUGE}
is truncated at some order in the derivatives, but it is clear that
\eqref{eq:Moyal-commutation-non-commutative} is correct, since these
higher orders will never appear in higher derivatives of $p_{\alpha}$
or $x^{\alpha}$. Many mathematical constructions start in practise
with supposing commutations relations of the form \eqref{eq:Moyal-commutation-non-commutative}.
Here they are the consequences of the Moyal product \eqref{eq:MOYAL-EXPANSION-GAUGE}.

\subsection{Quantum kinetic theory\label{subsec:Quantum-kinetic-theory}}

We now turn to the establishment of quantum kinetic theory using \eqref{eq:MOYAL-EXPANSION-GAUGE}.
The starting point of such constructions are the Dyson's equations
\begin{equation}
\int dy\int d\tau\left[G^{-1}\left(x_{1},t_{1};y,\tau\right)G\left(y,\tau;x_{2},t_{2}\right)\right]=\delta\left(x_{1}-x_{2}\right)\delta\left(t_{1}-t_{2}\right)\label{eq:Dyson-left}
\end{equation}
and 
\begin{equation}
\int dy\int d\tau\left[G\left(x_{1},t_{1};y,\tau\right)G^{-1}\left(y,\tau;x_{2},t_{2}\right)\right]=\delta\left(x_{1}-x_{2}\right)\delta\left(t_{1}-t_{2}\right)\label{eq:Dyson-right}
\end{equation}
from which one takes the sum and difference of the Wigner transforms.
When the inverse Green's function $G^{-1}$ presents a gauge structure,
it is natural to use the covariant Wigner transform \eqref{eq:gauge-covariant-Wigner-transform}
instead of the bare one \eqref{eq:Wigner}. Thus one would like to
apply the covariant Moyal expansion to the classic quantity $G^{-1}\left(p,x\right)$.

We first suppose time-independent problems, in which case it is natural
to write the quasi-classic Green's function (we aim at recovering
the results obtained in \citep{Bergeret2014} for superconducting
transport, in a simpler way) 
\begin{equation}
G^{-1}\left(p,x,t_{1},t_{2}\right)=\tau_{3}\left(\mathbf{i}\dfrac{\partial}{\partial t_{1}}+A_{0}\left(x\right)\right)-\left(\dfrac{p^{2}}{2m}-\mu\right)+\Delta\left(x\right)
\end{equation}
with the time-sector gauge-potential $A_{0}$ responsible for the
Zeeman splitting, and supposed to be space-dependent only (see section
\ref{subsec:Emergence-of-Gauge-structures}). The term $\Delta\left(x\right)$
can be either thought as a potential, or as the superconducting gap
in the mean-field approximation. The $\tau_{3}$-Pauli matrix is responsible
for the particle-hole redundancy in superconductors ($\Delta\propto\tau_{i}g_{i}\left(x\right)$
in that case), and can be taken as $\tau_{3}\rightarrow1$ in case
of non-superconducting systems. Then the Dyson's equation \eqref{eq:Dyson-left}
transforms to 
\begin{multline}
G^{-1}\left(p,x\right)\star G\left(p,x\right)=G^{-1}\left(p,x\right)G\left(p,x\right)+\dfrac{\mathbf{i}}{2}v_{i}\dfrac{\mathfrak{D}G}{\partial x^{i}}+\dfrac{\mathbf{i}}{2}\left(\dfrac{\mathfrak{D}\Delta}{\partial x^{i}}+\tau_{3}F_{i0}\right)\dfrac{\partial G}{\partial p_{i}}\\
+\dfrac{\mathbf{i}}{8}\left(3v_{i}F_{ij}\dfrac{\partial G}{\partial p_{j}}+\dfrac{\partial G}{\partial p_{j}}v_{i}F_{ij}\right)+\cdots
\end{multline}
where $\mathfrak{D}_{i}A_{0}=\partial_{i}A_{0}-\mathbf{i}\left[A_{i},A_{0}\right]\equiv F_{i0}\left(x\right)$
corresponds to the electric-like gauge-field in the pure static limit.
Doing the same manipulations for the right-applied Dyson's equation
\eqref{eq:Dyson-right} with the propagator
\begin{equation}
G^{-1}\left(p,x,t_{1},t_{2}\right)=\tau_{3}\left(-\mathbf{i}\dfrac{\partial^{\dagger}}{\partial t_{2}}+A_{0}\left(x\right)\right)-\left(\dfrac{p^{2}}{2m}-\mu\right)+\Delta\left(x\right)
\end{equation}
and taking the difference between the two obtained gradient expansions
finally leads to the so-called transport equation
\begin{multline}
\mathbf{i}\left(\tau_{3}\dfrac{\partial G}{\partial t_{1}}+\dfrac{\partial G}{\partial t_{2}}\tau_{3}\right)+\mathbf{i}v_{i}\dfrac{\mathfrak{D}G}{\partial x^{i}}+\left[\tau_{3}A_{0}\left(x\right)+\Delta\left(x\right),G\left(p,x,t_{1},t_{2}\right)\right]\\
+\dfrac{\mathbf{i}}{2}\left\{ \tau_{3}F_{0i}+v_{k}F_{ki},\dfrac{\partial G}{\partial p_{i}}\right\} +\dfrac{\mathbf{i}}{2}\left\{ \dfrac{\mathfrak{D}\Delta}{\partial x^{i}},\dfrac{\partial\mathbf{G}}{\partial p_{i}}\right\} =0\label{eq:Bergeret-Tokatly-transport}
\end{multline}
as obtained in \citep{Bergeret2014} after long algebra. In \citep{Bergeret2014},
the term $\mathfrak{D}_{i}\Delta$ is absent as being irrelevant for
physical reasons. Also, a self-energy term due to the possible impurities
was included in \citep{Bergeret2014} and has been discarded here,
for simplicity. We thus see that the Moyal expansion \eqref{eq:MOYAL-EXPANSION-GAUGE}
allows to find quite easilly the quasi-classic transport equation
of a classical action $G^{-1}\left(p,x\right)$ in a much simpler
way than the direct evaluation of the covariant Wigner transform \eqref{eq:gauge-covariant-Wigner-transform}.

In \eqref{eq:Bergeret-Tokatly-transport}, the time dependency was
not transformed into the time-frequency representation of the Wigner
transform. One can nevertheless suppose some quadri-vectors $p_{\alpha}\equiv\left(\omega,p_{i}\right)$
and $x^{\alpha}=\left(-t,x^{i}\right)$ such that 
\begin{equation}
\dfrac{\partial O_{1}}{\partial x^{\alpha}}\dfrac{\partial O_{2}}{\partial p_{\alpha}}=\dfrac{\partial O_{1}}{\partial x_{i}}\dfrac{\partial O_{2}}{\partial p_{i}}-\dfrac{\partial O_{1}}{\partial t}\dfrac{\partial O_{2}}{\partial\omega}
\end{equation}
in \eqref{eq:MOYAL-EXPANSION-GAUGE}, see section \ref{subsec:Conventions}
for more details. Then the gauge-covariant Wigner transform of the
time-derivative $D_{0}=\partial/\partial t-\mathbf{i}A_{0}$ corresponds
to $\omega$, according to the recipe of section \ref{subsec:Non-commutative-dressing}.
Starting then from 
\begin{equation}
G^{-1}\left(p,\omega,x,t\right)=\tau_{3}\omega-\left(\dfrac{p^{2}}{2m}-\mu\right)+\Delta\left(x,t\right)\label{eq:G-1-omega-t-p-x}
\end{equation}
and applying the Moyal rule \eqref{eq:MOYAL-EXPANSION-GAUGE} to the
associated Dyson equations \eqref{eq:Dyson-left} applied to the left
and \eqref{eq:Dyson-right} applied to the right (the propagator \eqref{eq:G-1-omega-t-p-x}
is the same in both positions in the classic representation), then
taking the difference between the two obtained gradient expansions,
one gets 
\begin{multline}
\dfrac{\mathbf{i}}{2}\left\{ \tau_{3},\dfrac{\mathfrak{D}G}{\partial t}\right\} +\mathbf{i}v_{i}\dfrac{\mathfrak{D}G}{\partial x^{i}}+\left[\tau_{3}\omega+\Delta\left(x,t\right),G\left(p,x,\omega,t\right)\right]\\
-\dfrac{\mathbf{i}}{8}\left(F_{0j}\left(3\tau_{3}\dfrac{\partial G}{\partial p_{j}}+\dfrac{\partial G}{\partial p_{j}}\tau_{3}\right)+\left(\tau_{3}\dfrac{\partial G}{\partial p_{j}}+3\dfrac{\partial G}{\partial p_{j}}\tau_{3}\right)F_{0j}\right)\\
+\dfrac{\mathbf{i}}{2}\left\{ v_{i}F_{ij},\dfrac{\partial G}{\partial p_{j}}\right\} +\dfrac{\mathbf{i}}{2}\left\{ v_{i}F_{i0},\dfrac{\partial G}{\partial\omega}\right\} +\dfrac{\mathbf{i}}{2}\left\{ \dfrac{\mathfrak{D}\Delta}{\partial x^{i}},\dfrac{\partial G}{\partial p_{i}}\right\} -\dfrac{\mathbf{i}}{2}\left\{ \dfrac{\mathfrak{D}\Delta}{\partial t},\dfrac{\partial G}{\partial\omega}\right\} =0\label{eq:TRANSPORT-PHASE-SPACE-TIME}
\end{multline}
for the transport equation of a singlet superconductor with spin-orbit
(formally written as a gauge-potential $A_{i}$) and spin-splitting
(a Zeeman term in the form of a time-sector gauge-potential $A_{0}$)
couplings (see section \ref{subsec:Emergence-of-Gauge-structures}).
This transport equation was first obtained in \citep{Konschelle2014},
and we can appreciate the concision of the present approach using
the Moyal expansion \eqref{eq:MOYAL-EXPANSION-GAUGE} with the cumbersome
derivation proposed in \citep{Konschelle2014}. The limits of a normal
metal in the presence of non-Abelian gauge fields (when $\tau_{3}\rightarrow1$
and $\Delta\rightarrow0$, see \citep{Gorini2010}) and of a superconductor
with electromagnetic coupling (i.e. an Abelian gauge-field, see \citep{Kopnin1994})
can be obtained from such an equation. An other equation, given by
the sum of the two Dyson's equations once Wigner transformed, is also
given in \citep{Konschelle2014}, and can be obtained as well from
the Moyal expansion \eqref{eq:MOYAL-EXPANSION-GAUGE} instead of the
long derivation done in \citep{Konschelle2014}.

We just seen in this section that the Moyal expansion \eqref{eq:MOYAL-EXPANSION-GAUGE}
provides a convenient way to overcome the complicated calculations
done in the establishments and justifications of gauge-covariant transport
equations. We here reviewed the condensed matter situations of such
equations, where the above equations (in one form or another) have
been used to establish many results, especially in the field of magneto-electric
effects for low-energy spin manipulation using the macroscopic coherence
of superconductivity, the so-called super-spintronics \citep{Linder2015,Eschrig2015a,Aprili2017}.
Reader interested in the recent results might consult the review \citep{Raimondi2016}
for normal metallic systems and \citep{Bergeret2014a,Konschelle2015,Jacobsen2015,Reeg2015,Konschelle2016,Konschelle2016a,Espedal2016,Malshukov2016,Bergeret2016,Reeg2017}
for some descriptions of super-spintronics effects. Similar calculations
can be done for the quark-gluon plasma, using eventually the Wigner
function (i.e. the Wigner transform of the density operator) instead
of the Green's function \citep{Elze1989,Weigert1991,Wong1996}. There
as well, to know the Moyal expansion \eqref{eq:MOYAL-EXPANSION-GAUGE}
would have drastically simplified the obtention of the transport equations.
In particular, it would simplify the search for effective actions
at the one-loop order, see e.g. \citep{Pletnev1999,Pletnev2001,Salcedo2007}
and references therein.

\subsection{Topology in the momentum space\label{subsec:Topology-in-the-momentum-space}}

In addition to the establishment of transport equations reviewed in
section \ref{sec:Moyal-product-in-action}, the gradient expansion
\eqref{eq:MOYAL-EXPANSION-GAUGE} allows to evaluate perturbatively
the Green's function of a given system. Then it becomes possible,
without much assumptions (like linear-response theory for systems
slightly out of equilibrium for instance), to establish generic properties
of matter using quasi-classic methods. This is what we review quickly
in this section. As a simple yet interesting example, the momentum-space
topology introduced by Volovik \citep{Volovik2003}, is reviewed.
In the way, we will generalise the results previously known for Abelian
gauge-theory to non-Abelian situations at no cost. Indeed, the situations
described below are usually identified using the flat-space Moyal
expansion \eqref{eq:Moyal-expansion}, whereas we establish these
results using the gauge-covariant Moyal expansion \eqref{eq:MOYAL-EXPANSION-GAUGE}
instead.

The basic idea of the obtention of topological quantities is to use
the Dyson's equation \eqref{eq:Dyson-left} in the phase-space, 
\begin{equation}
G^{-1}\left(p,x\right)\star G\left(p,x\right)=1
\end{equation}
and to expand the solution for $G\left(p,x\right)$ in the form $G=G_{0}\left(1+G_{2}+\cdots\right)$
in power of gradient, once replacing the star product with \eqref{eq:MOYAL-EXPANSION-GAUGE}.
One obtains in that way
\begin{equation}
G_{0}\left(p,x\right)=\left[G^{-1}\left(p,x\right)\right]^{-1}\label{eq:G0-gradient}
\end{equation}
at zero-th order in gradient. Thus the zero-th order Green's function
is just the classical Green's function. There is no first order term,
and the second one reads
\begin{multline}
G_{2}\left(p,x\right)=-\dfrac{\mathbf{i}}{2}\left(\dfrac{\mathfrak{D}G_{0}^{-1}}{\partial x^{\alpha}}\dfrac{\partial G_{0}}{\partial p_{\alpha}}-\dfrac{\partial G_{0}^{-1}}{\partial p_{\alpha}}\dfrac{\mathfrak{D}G_{0}}{\partial x^{\alpha}}\right)+\\
+\dfrac{\mathbf{i}}{8}\left(F_{\alpha\beta}\left(x\right)\dfrac{\partial G_{0}^{-1}}{\partial p_{\alpha}}\dfrac{\partial G_{0}}{\partial p_{\beta}}+2\dfrac{\partial G_{0}^{-1}}{\partial p_{\alpha}}F_{\alpha\beta}\left(x\right)\dfrac{\partial G_{0}}{\partial p_{\beta}}+\dfrac{\partial G_{0}^{-1}}{\partial p_{\alpha}}\dfrac{\partial G_{0}}{\partial p_{\beta}}F_{\alpha\beta}\left(x\right)\right)\label{eq:G2-gradient}
\end{multline}
which, to the best of our knowledge, is a novel result. We stop at
this order to recover known results in the case of normal metal (where
the Chern number appears) and superfluids (where the Pontryagin number
appears). Higher order terms will be inspected in subsequent studies.

The strategy is for the moment to inject the perturbative Green's
function into the definition of the current (see \citep{Zubkov2016}
for a justification -- the notations are relativistic with $j^{\alpha}\equiv\left(\rho,j_{i}\right)$
a notation for the density of charge and the current vector, and $p_{\alpha}\equiv\left(\omega,\boldsymbol{p}\right)$
in the reciprocal space-time of frequency and momentum) 
\begin{equation}
j^{\alpha}=\dfrac{\mathbf{i}}{\deg}\dfrac{e}{4}\int\dfrac{d\omega}{2\pi}\int\dfrac{dp}{2\pi}\Tr\left\{ \dfrac{\partial G_{0}^{-1}}{\partial p_{\alpha}}G\left(p,x\right)\right\} \label{eq:j-DEF}
\end{equation}
to get $j^{\alpha}=j_{\left(1\right)}^{\alpha}+j_{\left(2\right)}^{\alpha}+\cdots$
with 
\begin{equation}
j_{\left(1\right)}^{\alpha}=\dfrac{1}{\deg}\dfrac{e}{8}\int\dfrac{d\omega}{2\pi}\int\dfrac{dp}{2\pi}\Tr\left\{ \dfrac{\mathfrak{D}G_{0}^{-1}}{\partial x^{\beta}}G_{0}\left[\dfrac{\partial G_{0}^{-1}}{\partial p_{\alpha}}G_{0},\dfrac{\partial G_{0}^{-1}}{\partial p_{\beta}}G_{0}\right]\right\} 
\end{equation}
\begin{equation}
j_{\left(2\right)}^{\alpha}=\dfrac{1}{\deg}\dfrac{e}{8}\int\dfrac{d\omega}{2\pi}\int\dfrac{dp}{2\pi}\Tr\left\{ F_{\beta\gamma}\left(x\right)\dfrac{\partial G_{0}^{-1}}{\partial p_{\alpha}}G_{0}\dfrac{\partial G_{0}^{-1}}{\partial p_{\beta}}G_{0}\dfrac{\partial G_{0}^{-1}}{\partial p_{\gamma}}G_{0}\right\} \label{eq:j-2}
\end{equation}
up to higher order in the gradients. We supposed that $\left[F_{\alpha\beta},G_{0}\right]=0$
since usually the Green's function is the bare one, as we discuss
in section \ref{sec:Moyal-product-in-action}, see especially \eqref{eq:G-1-omega-t-p-x}
for the example of a superconductor. In \eqref{eq:j-DEF}, we defined
the degeneracy number $\deg$ as the number of redundant degrees of
freedom of the Green's function introduced in order to account for
the symmetry of the model. It is essentially in use when dealing with
both normal and superconducting systems. In this later case, $\deg=2$
accounts for the particle-hole symmetry of the BCS model, whereas
$\deg=1$ in normal systems.

Let us have a look at $j_{\left(1\right)}^{\alpha}$. When the covariant
derivative is replaced by the bare one $\mathfrak{D}\rightarrow\partial$,
this term is the only contribution coming from the $\star_{0}$-expansion
\eqref{eq:Moyal-expansion} in the flat space. In particular $j_{\left(2\right)}=0$
in that case. There is a trick to get the linear-response current
in the case of Abelian fields, even when one starts from the $\star_{0}$-product
instead of the covariant $\star$-one \citep{Zubkov2016,Mera2017}.
If one supposes that the position dependency of the classical Green's
function reads $G_{0}\left(p-A\left(x\right)\right)$ -- the minimal
substitution recipe -- the expansions $G_{0}\left(p-A\right)=G_{0}\left(p\right)-A_{\alpha}\left(x\right)\partial G_{0}/\partial p_{\alpha}+\cdots$
and a similar one for $G_{0}^{-1}$ are allowed and one gets
\begin{equation}
j_{\left(1\right),\text{min}}^{\alpha}=j_{\left(1\right)}^{\alpha}\left(\mathfrak{D}\rightarrow\partial\right)+\dfrac{e}{8\deg}\int\dfrac{d\omega}{2\pi}\int\dfrac{dp}{2\pi}\Tr\left\{ F_{\beta\gamma}^{\text{Ab.}}\left(x\right)\dfrac{\partial G_{0}^{-1}}{\partial p_{\alpha}}G_{0}\dfrac{\partial G_{0}^{-1}}{\partial p_{\beta}}G_{0}\dfrac{\partial G_{0}^{-1}}{\partial p_{\gamma}}G_{0}\right\} \label{eq:j-1-min}
\end{equation}
with the Abelian gauge field $F_{\alpha\beta}^{\text{Ab.}}=\partial_{\alpha}A_{\beta}-\partial_{\beta}A_{\alpha}$,
see details in \citep{Zubkov2016,Mera2017}. The extra contribution
is exactly $j_{\left(2\right)}$ whenever the gauge fields are Abelian.
Despite giving similar contributions at the leading order, the strategy
outlined in this paragraph is not well justified, because the star
product $\star_{0}$ does not guarantee to obtain only covariant terms.
In addition, the above heuristic minimal substitution to get $j_{\left(1\right),\text{min}}$
seems to be hardly justified in the case of non-Abelian fields, whereas
our strategy here to dress the bare Green's function $G_{0}\left(p,x\right)$
by the gauge field using the covariant $\star$-product expansion
\eqref{eq:MOYAL-EXPANSION-GAUGE} gives a straightforward evaluation
of the current in the linear-response framework, either in the Abelian
or the non-Abelian cases. In addition, we impose no hypothesis about
the structure of $G_{0}\left(p,x\right)$ when we derived \eqref{eq:j-2}.

We come back to the covariant calculation, and pursue our interpretation
of the term $j_{\left(2\right)}$ obtained in \eqref{eq:j-2}. Due
to the antisymmetry of the gauge-field, and the cyclic property of
the trace, one can define the symbol 
\begin{equation}
R^{\alpha\beta\gamma}=\dfrac{e}{8}\int\dfrac{d\omega}{2\pi}\int\dfrac{dp}{2\pi}\Tr\left\{ \dfrac{\partial G_{0}^{-1}}{\partial p_{\alpha}}G_{0}\dfrac{\partial G_{0}^{-1}}{\partial p_{\beta}}G_{0}\dfrac{\partial G_{0}^{-1}}{\partial p_{\gamma}}G_{0}\right\} 
\end{equation}
as being totally antisymmetric in its three indices. In the case of
Abelian gauge field, one can represent \eqref{eq:j-2}, with $\dim$
the space dimension of the problem, as (notations taken from \citep{Zubkov2016})
\begin{equation}
j_{\left(2\right)}^{\alpha}\left(x\right)=R^{\alpha\beta\gamma}\left(x\right)F_{\beta\gamma}\left(x\right)=\dfrac{1}{\deg}\dfrac{e}{4}\begin{cases}
\dfrac{\varepsilon^{\alpha\beta\gamma}}{4\pi}F_{\beta\gamma}\mathcal{M}, & \dim=2\\
\dfrac{\varepsilon^{\alpha\beta\gamma\delta}}{4\pi^{2}}F_{\beta\gamma}N_{\delta}, & \dim=3
\end{cases}
\end{equation}
where we define the quantities 
\begin{equation}
\mathcal{M}=-\dfrac{\varepsilon_{\alpha\beta\gamma}}{3!4\pi^{2}}\iiint d\omega dp_{x}dp_{y}\Tr\left\{ \dfrac{\partial G_{0}^{-1}}{\partial p_{\alpha}}G_{0}\dfrac{\partial G_{0}^{-1}}{\partial p_{\beta}}G_{0}\dfrac{\partial G_{0}^{-1}}{\partial p_{\gamma}}G_{0}\right\} \label{eq:M-definition}
\end{equation}
\begin{equation}
N_{\delta}=-\dfrac{\varepsilon_{\alpha\beta\gamma\delta}}{3!8\pi^{2}}\iiiint d\omega dp_{x}dp_{y}dp_{z}\Tr\left\{ \dfrac{\partial G_{0}^{-1}}{\partial p_{\alpha}}G_{0}\dfrac{\partial G_{0}^{-1}}{\partial p_{\beta}}G_{0}\dfrac{\partial G_{0}^{-1}}{\partial p_{\gamma}}G_{0}\right\} 
\end{equation}
and one can further show that \eqref{eq:M-definition} reduces to
some topological numbers in peculiar circonstances. For instance,
in the case of a normal metal without interactions at zero temperature,
$\deg=1$ and the Green's function reduces to 
\begin{equation}
G_{0}\left(p,x\right)=\sum_{n}\dfrac{\left|n,p\right\rangle \left\langle n,p\right|}{\mathbf{i}\omega-E_{n}\left(p\right)}\label{eq:G-normal-metal}
\end{equation}
where the electronic states are represented by the wave-function $\left|n,p\right\rangle $,
with $n$ labeling the eigenenergies $E_{n}$. Injecting \eqref{eq:G-normal-metal}
in \eqref{eq:M-definition}, and performing the contour integration
over the imaginary Matsubara frequencies $\mathbf{i}\omega$ at zero
temperature, one can evaluate (see details in \citep{Zubkov2016,Mera2017}),
\begin{equation}
\mathcal{M}=\dfrac{1}{4\pi}\iint dp_{x}dp_{y}\left[\mathcal{F}_{xy}\right]\label{eq:M-Chern}
\end{equation}
where $\mathcal{A}_{i}=-\mathbf{i}\left\langle n\right|\partial_{i}\left|n\right\rangle $
and $\mathcal{F}_{ij}=\partial_{i}\mathcal{A}_{j}-\partial_{j}\mathcal{A}_{i}$
corresponds to the Berry connection and curvature, respectively. This
last expression is nothing but the celebrated quantisation formula
for the transconductance in 2D systems under magnetic field with $\mathcal{M}$
the Chern number \citep{Thouless1982,Simon1983,Niu1985}. 

We obtained this topological result in the case of a clean metal without
interaction, represented by the expression \eqref{eq:G-normal-metal}.
Nevertheless, variation of $G_{0}\rightarrow G_{0}+\delta G$ in the
definition \eqref{eq:M-definition} indicates that this quantity is
in fact invariant under small perturbation of the Green's function,
hence it gets some topological significations irrespective of the
explicit representation of the Green's function \citep{Grinevich1988,Volovik2003},
and might continue to carry some topological significations even in
the presence of interactions \citep{Gurarie2011,Essin2011}.

An other important example of the use of the gradient expansion is
given by the quantisation of particle transconductance in the case
of bi-dimensionnal Helium superfluids. In that later situation, $\deg=2$
and the Green's function reads 
\begin{equation}
G_{0}^{-1}\left(p,x\right)=\omega-\boldsymbol{\tau\cdot m}\left(p\right)\label{eq:G-superfluids}
\end{equation}
with $\tau_{i}$ the Pauli matrices which span the particule-hole
algebra of the Cooper condensation in the superfluid. The vector $\boldsymbol{m}$
encodes the symmetries of the superfluid gap. One then gets 
\begin{equation}
\mathcal{M}=-\dfrac{1}{16\pi}\iint dp_{x}dp_{y}\left[\boldsymbol{\hat{m}\cdot}\left(\dfrac{\partial\boldsymbol{\hat{m}}}{\partial p_{x}}\times\dfrac{\partial\boldsymbol{\hat{m}}}{\partial p_{y}}\right)\right]\label{eq:M-Pontryagin}
\end{equation}
which is nothing but the Pontryagin number. This quantity has been
derived in \citep{Volovik1988b,Volovik1989}, showing how the superfluids
phenomenology can be similar to the quantum Hall effect, exhibiting
transconductance quantisation, though of different origin. Indeed,
in the quantum Hall case, the quantisation is due to the non-trivial
electronic band structure under a magnetic field, whereas in superfluids,
the non-trivial topology orginates from the particle-hole space and
the symmetries of the superfluid gap, the Pauli-$\tau$ matrices in
\eqref{eq:G-superfluids}.

One more time, note that the results \eqref{eq:M-Chern} and \eqref{eq:M-Pontryagin}
were previously obtained using a non-covariant gradient expansion.
They have then been established using the trick explained around \eqref{eq:j-1-min},
using some minimal substitution $G_{0}\left(p-A\right)$ instead of
the bare Green's functions. In contrary, we established them using
the covariant expansion of the $\star$-product, and \eqref{eq:j-2}
is therefore more general than the tricks used in the existing literature
\citep{Volovik1988b,Volovik1989,Zubkov2016,Mera2017}. At least one
can show that our approach is gauge covariant to start with.

The above review helps us to understand the structure of the gauge-covariant
Moyal expansion \eqref{eq:MOYAL-EXPANSION-GAUGE}. Indeed, none of
the topological numbers $\mathcal{M}$ in \eqref{eq:M-Chern} or \eqref{eq:M-Pontryagin}
would have been obtained for a pure classic action \eqref{eq:G-normal-metal}
or \eqref{eq:G-superfluids}. It is because the Moyal expansion \eqref{eq:MOYAL-EXPANSION-GAUGE}
dresses the bare actions \eqref{eq:G-normal-metal} and \eqref{eq:G-superfluids}
by some gauge fields that the topological obstructions \eqref{eq:M-Chern}
and \eqref{eq:M-Pontryagin} can be extracted from it. Though the
gauge structure in this section explicitly comes from the Abelian
electromagnetic structure, its origin in the general case of \eqref{eq:MOYAL-EXPANSION-GAUGE}
might appear unclear at first sight. In fact, we never imposed any
restriction on the gauge field, beyond its generic definition \eqref{eq:F-DEFINITION}.
So, prior to any gradient expansion, the gauge structure of the theory
must be extracted from extra considerations, like e.g. microscopic
theories. In fact, one usually starts from a microscopic theory, from
which the gauge structure can be established on the basis of symmetry
construction. Once the gauge structure is established in the form
of the minimal substitution $p\rightarrow p-A$, one can apply the
gauge-covariant Wigner transform \eqref{eq:gauge-covariant-Wigner-transform}
to the equation of motion (e.g. the Schrödinger or the Dyson equation)
on the bare theory, such that the gauge-covariant gradient expansion
\eqref{eq:MOYAL-EXPANSION-GAUGE} will dress the perturbative theory
by gauge interactions. At this level the gauge fields are classical
fields, and the expansion \eqref{eq:MOYAL-EXPANSION-GAUGE} can be
seen as a perturbative expansion in both gradients and gauge fields
(higher orders in the gauge field will naturally appear in the higher
orders, see section \ref{sec:Gauge-covariant-Moyal-star-product}
for more details). A few strategies to understand how the gauge structure
emerges in microscopic theories have been given in section \ref{subsec:Emergence-of-Gauge-structures}.

To conclude this section, let us calculate the spin current in 2D
(the definition follows straightforwardly from arguments in \citep{Zubkov2016}
and \citep{Konschelle2014}, see also \citep{Konschelle2016a} for
a similar definition) 
\begin{align}
J_{a}^{\alpha}\left(x\right) & =\dfrac{\mathbf{i}}{4\deg}\int\dfrac{d\omega}{2\pi}\int\dfrac{dp_{x}}{2\pi}\int\dfrac{dp_{y}}{2\pi}\Tr\left\{ \sigma^{a}\dfrac{\partial G_{0}^{-1}}{\partial p_{\alpha}}G_{0}\left(p,x\right)\right\} \nonumber \\
 & =\dfrac{1}{8\deg}\int\dfrac{d\omega}{2\pi}\int\dfrac{dp_{x}}{2\pi}\int\dfrac{dp_{y}}{2\pi}\Tr\left\{ \sigma^{a}F_{\beta\gamma}\left(x\right)\dfrac{\partial G_{0}^{-1}}{\partial p_{\alpha}}G_{0}\dfrac{\partial G_{0}^{-1}}{\partial p_{\beta}}G_{0}\dfrac{\partial G_{0}^{-1}}{\partial p_{\gamma}}G_{0}\right\} \nonumber \\
 & =\dfrac{1}{4\deg}\dfrac{\varepsilon^{\alpha\beta\gamma}}{4\pi}F_{\beta\gamma}^{a}\left(x\right)\mathcal{M}\label{eq:spin-current-M}
\end{align}
where we supposed the Green's function $G_{0}$ to be independent
of the spin, and we expand the non-Abelian gauge field $F_{\alpha\beta}\equiv F_{\alpha\beta}^{a}\sigma^{a}/2$
when it describes the spin degree of freedom with Pauli algebra span
by spin matrices $\sigma^{a}$, $a\in\left\{ 1,2,3\right\} $. Then
the trace applies on the spin degree of freedom only, and project
the gauge field toward the Pauli matrix $\sigma^{a}$ present in the
definition of the spin current. 

The spin-current in \eqref{eq:spin-current-M} can then be generated
by pure spin-orbit interactions (like Rashba or Dresselhaus ones,
see \citep{Winkler2003,Berche2013}), and might be quantised in the
limit of zero temperature, following the same arguments giving \eqref{eq:M-Chern}
in normal metal or \eqref{eq:M-Pontryagin} in superfluid/superconducting
systems. Nevertheless, the spin current is impossible to measure,
since it contains 5 dimensions (three for the spin $a\in\left\{ 1,2,3\right\} $
and two for space $\alpha\in\left\{ 1,2\right\} $) in our three dimensional
experimental world. An other way to understand why the spin current
can not be measured is to realise that it is not conserved in the
usual sense (namely as $\partial_{\alpha}j^{\alpha}=0$), but in a
non-Abelian gauge covariant way (namely $\mathfrak{D}_{\alpha}J^{\alpha}=\partial_{\alpha}J^{\alpha}-\mathbf{i}\left[A_{\alpha},J^{\alpha}\right]=0$;
the part $\mathbf{i}\left[A_{\alpha},J^{\alpha}\right]$ is usually
considered as a spin-torque), see e.g. \citep{Jin2006,Tokatly2017}
and references therein.

Many other consequences of the gauge-covariant gradient expansion
\eqref{eq:MOYAL-EXPANSION-GAUGE} are still to come, and will be the
subject of subsequent works. At the moment, we turn to the presentation
of the method leading to the expansion \eqref{eq:MOYAL-EXPANSION-GAUGE}. 

\section{Gauge-covariant Moyal star-product\label{sec:Gauge-covariant-Moyal-star-product}}

We now attack the heart section of this study. We will proove the
gradient expansion \eqref{eq:MOYAL-EXPANSION-GAUGE}, and add the
next order term to it, namely the term of fourth order in gradient.
A generic method will be presented, based in the difference between
the gauge-covariant Wigner transform \eqref{eq:gauge-covariant-Wigner-transform}
and a gauge-fixed one in the Fock-Schwinger gauge (also called Poincaré-relativistic,
or axial or radial gauge) proposed in \citep{Weigert1991}. Section
\ref{subsec:Two-different-Wigner-transforms} provides a self-contained
justification of the gauge-covariant and the gauge-fixed Wigner transforms.
The method consists in realizing that in the Fock-Schwinger gauge,
the gauge-covariant transform \eqref{eq:gauge-covariant-Wigner-transform}
looks formally like the Wigner transform \eqref{eq:Wigner} in the
flat space. Then one can apply the well-known result \eqref{eq:Moyal-expansion}
to it. The mapping from the gauge-flat to gauge-covariant formulation
can be done by promoting the successive $x$-derivatives to some covariant
derivatives, including gauge fields corrections. This is the task
accomplished in section \ref{subsec:Substitutions-for-the-derivatives}.
Unfortunately, I am not aware of any trick allowing the susbtitution
to be done at any order, and so the substitution is done order by
order. This is where the pedestrian approach appears. Section \ref{subsec:Substitutions-for-the-derivatives}
is quite technical, and might be skipped in first reading since it
does not participate in the understanding of the rest of the presentation.
The gauge-covariant gradient expansion is then given in section \ref{subsec:Gauge-covariant-gradient-expansion}.
The two limiting cases of Abelian gauge fields and pure gauge problems
are given in section \ref{subsec:Abelian-and-pure-gauge}.

\subsection{Two different Wigner transforms\label{subsec:Two-different-Wigner-transforms}}

We begin by a short review of the gauge-covariant Wigner transform
proposed in \citep{Winter1984,Elze1986,Elze1986a} and the gauge-fixed
Wigner transform proposed in \citep{Weigert1991}.

The story starts with the following problem. Given the Wigner transform
\eqref{eq:Wigner}, and the possibility for the operators $O\left(x_{1},x_{2}\right)$
to change under a gauge transformation to $O^{\prime}\left(x_{1},x_{2}\right)=R\left(x_{1}\right)O\left(x_{1},x_{2}\right)R^{-1}\left(x_{2}\right)$,
how may we construct a Wigner transform $O\left(p,x\right)$ which
would be gauge-covariant ? Namely we would like $O\left(p,x\right)$
to transform as $O^{\prime}\left(p,x\right)=R\left(x\right)O\left(p,x\right)R^{-1}\left(x\right)$
under a gauge transformation. The solution to this problem, first
proposed in \citep{Luttinger1951,Stratonovich1956}, is to realise
that \eqref{eq:Wigner} can be writen in the formal representation
\begin{equation}
O\left(p,x\right)=\int d\mathfrak{x}\left[e^{-\mathbf{i}p\cdot\mathfrak{x}}\left(e^{\mathfrak{x}\cdot\partial_{x_{1}}/2}O\left(x_{1},x_{2}\right)e^{-\mathfrak{x}\cdot\partial_{x_{2}}^{\dagger}/2}\right)_{x_{1}=x_{2}=x}\right]\label{eq:Wigner-formal-dx}
\end{equation}
where the dagger-derivative $O\partial_{x}^{\dagger}\equiv\partial_{x}O$
applies to the left. The exponentiated derivative shifts the argument
of any function of its argument, $e^{\mathfrak{x}\cdot\partial_{x}}f\left(x\right)e^{-\mathfrak{x}\cdot\partial_{x}}=f\left(x+\mathfrak{x}\right)$.
In effect, the formal writing \eqref{eq:Wigner-formal-dx} is strictly
equivalent to the previous definition \eqref{eq:Wigner}. 

Under the representation \eqref{eq:Wigner-formal-dx}, it is straightforward
to generalise the Wigner transform to a covariant representation.
It is sufficient to replace the derivative by its covariant version
$\partial_{x}\rightarrow D_{x}=\partial_{x}-\mathbf{i}A_{x}\left(x\right)$
and $\partial_{x}^{\dagger}\rightarrow D_{x}^{\dagger}=\partial_{x}^{\dagger}+\mathbf{i}A_{x}\left(x\right)$.
One obtains the representation \eqref{eq:gauge-covariant-Wigner-transform}
once we realise that
\begin{equation}
e^{\mathfrak{x}\cdot D_{x}}f\left(x\right)=U\left(x,x+\mathfrak{x}\right)f\left(x+\mathfrak{x}\right)\label{eq:U-exp-DEF}
\end{equation}
with the parallel transport operator given in \eqref{eq:parallel-transport}.
We can drop the index $x_{1}=x_{2}=x$ from now on in \eqref{eq:Wigner-formal-dx}.
This operator has the property of a Dyson's operator: $U\left(x,x\right)=1$,
$U\left(x_{1},y\right)U\left(y,x_{2}\right)=U\left(x_{1},x_{2}\right)$
and so $U\left(x_{1},x_{2}\right)$ is the inverse of $U\left(x_{2},x_{1}\right)$.
Details of the calculation done in \citep{Winter1984,Elze1986}, partially
reproduced in \citep{Konschelle2014}, show that the path connecting
the points $x$ and $x+\mathfrak{x}$ using the operator $e^{\mathfrak{x}\cdot D_{x}}$
must be a straight line.

For the following it is important to realise that the derivatives
of $U$ are not trivial. In fact, one has 
\begin{equation}
\dfrac{DU\left(b,a\right)}{\partial b^{\alpha}}U\left(a,b\right)=\mathbf{i}\left(b-a\right)^{\beta}\int_{0}^{1}ds\left[sU\left(b,\tau_{s}\right)F_{\beta\alpha}\left(\tau_{s}\right)U\left(\tau_{s},b\right)\right]\label{eq:DU}
\end{equation}
where the gauge field is defined as $F_{\alpha\beta}=\partial_{\alpha}A_{\beta}-\partial_{\beta}A_{\alpha}-\mathbf{i}\left[A_{\alpha},A_{\beta}\right]$
(with the convention $\partial_{\alpha}A_{\beta}\equiv\partial A_{\beta}/\partial x^{\alpha}$)
in the non-Abelian case, and $\tau_{s}=a+\left(b-a\right)s$ represents
the straight line between $b$ and $a$. Using the properties of the
parallel transport operation, one gets similarly 
\begin{equation}
U\left(a,b\right)\dfrac{U\left(b,a\right)D^{\dagger}}{\partial a^{\alpha}}=\mathbf{i}\left(b-a\right)^{\beta}\int_{0}^{1}ds\left[\left(1-s\right)U\left(a,\tau_{s}\right)F_{\beta\alpha}\left(\tau_{s}\right)U\left(\tau_{s},a\right)\right]
\end{equation}
for the derivative applied to the second variable. The presence of
$\left(1-s\right)$ ensures that the path is taken in the opposite
way.

The Wigner transform \eqref{eq:gauge-covariant-Wigner-transform}
is covariant with respect to the gauge transformation: $O^{\prime}\left(p,x\right)=R\left(x\right)O\left(p,x\right)R^{-1}\left(x\right)$,
as required. One can in particular applies the transformation 
\begin{multline}
O^{\left(Z\right)}\left(p,x\right)=U\left(Z,x\right)O\left(p,x\right)U\left(x,Z\right)\\
=\int d\mathfrak{x}\left[e^{-\mathbf{i}p\mathfrak{x}/\hbar}U\left(Z,x+\dfrac{\mathfrak{x}}{2}\right)O\left(x+\dfrac{\mathfrak{x}}{2},x-\dfrac{\mathfrak{x}}{2}\right)U\left(x-\dfrac{\mathfrak{x}}{2},Z\right)\right]\label{eq:OZ}
\end{multline}
where the external point $Z$ is arbitrary for the moment. Nevertheless,
it is the same parallel transport operator \eqref{eq:parallel-transport}
which has been used in the definition of $O^{\left(Z\right)}\left(p,x\right)$.
In particular, $U\left(Z,x\right)$ verifies \eqref{eq:DU}. Suppose
then that one applies the same gauge transformation to the gauge-potential,
defining 
\begin{align}
A_{\alpha}^{\left(Z\right)}\left(x\right) & =U\left(Z,x\right)A_{\alpha}\left(x\right)U\left(x,Z\right)+\mathbf{i}U\left(Z,x\right)\dfrac{\partial U\left(x,Z\right)}{\partial x^{\alpha}}\nonumber \\
 & =\left(Z-x\right)^{\beta}\int_{0}^{1}ds\left[sU\left(Z,\tau_{s}\right)F_{\beta\alpha}\left(\tau_{s}\right)U\left(\tau_{s},Z\right)\right]
\end{align}
with $\tau_{s}=Z+\left(x-Z\right)s$ for the path between $x$ and
$Z$, using \eqref{eq:DU} to kill $A_{\alpha}$. In particular, we
have, in this new gauge 
\begin{equation}
\left(x-Z\right)^{\alpha}A_{\alpha}^{\left(Z\right)}\left(x\right)=0
\end{equation}
since the gauge field is an antisymmetric quantity $F_{\alpha\beta}=-F_{\beta\alpha}$
by definition. This is nothing but the definition of the radial gauge,
hence the transformation \eqref{eq:OZ} is called the radial-gauge
Wigner transform, or gauge-fixed Wigner transform. In addition, one
has $A_{\alpha}^{\left(Z\right)}\left(x=Z\right)=0$ and so the covariant
derivatives reduce to the trivial ones at the point $x=Z$. This property
allows easy manipulations. Indeed, taking $x=Z$ in the definition
of the Wigner transform \eqref{eq:gauge-covariant-Wigner-transform}
but keeping $x_{1,2}=x\pm\mathfrak{x}/2$ as independent variables
reduces $O\left(p,x\right)$ to $O^{\left(Z\right)}\left(p,x\right)$
in \eqref{eq:OZ}. So the equation of motion for $O^{\left(Z\right)}\left(p,x\right)$
presents only flat-space derivatives $\partial_{x}$, and in particular,
the Moyal product for two operators $O_{1}^{\left(Z\right)}\left(p,x\right)$
and $O_{2}^{\left(Z\right)}\left(p,x\right)$ is the well known one
\eqref{eq:Moyal-expansion} since there is vanishing connection in
this gauge when $x=Z$. We thus have 
\begin{multline}
\int d\mathfrak{x}\int dy\left[e^{-\mathbf{i}p\mathfrak{x}/\hbar}U\left(Z,x+\dfrac{\mathfrak{x}}{2}\right)O_{1}\left(x+\dfrac{\mathfrak{x}}{2},y\right)O_{2}\left(y,x-\dfrac{\mathfrak{x}}{2}\right)U\left(x-\dfrac{\mathfrak{x}}{2},Z\right)\right]=\\
=O_{1}^{\left(Z\right)}\left(p,x\right)\star_{0}O_{2}^{\left(Z\right)}\left(p,x\right)=O_{1}^{\left(Z\right)}O_{2}^{\left(Z\right)}+\dfrac{\mathbf{i}}{2}\left(\dfrac{\partial O_{1}^{\left(Z\right)}}{\partial x^{\alpha}}\dfrac{\partial O_{2}^{\left(Z\right)}}{\partial p_{\alpha}}-\dfrac{\partial O_{1}^{\left(Z\right)}}{\partial p_{\alpha}}\dfrac{\partial O_{2}^{\left(Z\right)}}{\partial x^{\alpha}}\right)+\cdots\label{eq:Moyal-Z}
\end{multline}
at first order of the star product \eqref{eq:Moyal-expansion}. We
no more need to precise the variables $\left(p,x\right)$ for $O_{1,2}^{\left(Z\right)}$
since the gauge exponents specifies that we work in the phase space,
allowing simpler notations. 

We have thus established the Moyal product in the radial gauge: it
is just the usual Moyal $\star_{0}$-product \eqref{eq:Moyal-expansion}.
This is nevertheless not gauge covariant. To know the gauge-covariant
gradient expansion, one must perform the parallel transport in the
reverse sense, i.e. from the point $Z$ back to $x$. We thus take
the limit $Z\rightarrow x$ in the gradient expansion \eqref{eq:Moyal-Z}.
It consists in replacing $O_{1,2}^{\left(Z\right)}\rightarrow O_{1,2}\left(p,x\right)$
naively. Nevertheless, the space derivatives in \eqref{eq:Moyal-Z}
must be taken with care. One indeed realises that \citep{Weigert1991}
\begin{equation}
\dfrac{\partial O\left(p,x\right)}{\partial x^{\alpha}}=\left.\left(\dfrac{\partial}{\partial x^{\alpha}}+\dfrac{\partial}{\partial Z^{\alpha}}\right)O^{\left(Z\right)}\right|_{Z=x}\label{eq:derivative-substitution-def}
\end{equation}
by a direct comparison of the definitions of the Wigner transforms
in the radial gauge \eqref{eq:OZ} and in the covariant representation
\eqref{eq:gauge-covariant-Wigner-transform}. We finally have to explicitly
calculate the limit 
\begin{equation}
O_{1}\left(p,x\right)\star O_{2}\left(p,x\right)=\lim_{Z\rightarrow x}\left[O_{1}^{\left(Z\right)}\left(p,x\right)\star_{0}O_{2}^{\left(Z\right)}\left(p,x\right)\right]\label{eq:mapping-flat-covariant}
\end{equation}
using the substitution \eqref{eq:derivative-substitution-def}, e.g.
\begin{equation}
O_{1}\left(p,x\right)\star O_{2}\left(p,x\right)=O_{1}\left(p,x\right)O_{2}\left(p,x\right)+\dfrac{\mathbf{i}}{2}\left(\left.\dfrac{\partial O_{1}^{\left(Z\right)}}{\partial x^{\alpha}}\right|_{Z=x}\dfrac{\partial O_{2}}{\partial p_{\alpha}}-\dfrac{\partial O_{1}}{\partial p_{\alpha}}\left.\dfrac{\partial O_{2}^{\left(Z\right)}}{\partial x^{\alpha}}\right|_{Z=x}\right)+\cdots
\end{equation}
at the leading order. The momentum derivative do not make any trouble,
and we replace them straightforwardly in the above expression. The
$x$-derivative of $O^{\left(Z\right)}$ naturally appears in the
$\star_{0}$-product, so we just have to evaluate $\partial_{Z}O^{\left(Z\right)}$
and take its limit $Z\rightarrow x$. This can be done exactly using
the property \eqref{eq:DU} of the parallel transport, which let the
gauge field appearing. 

This is all for the general method, we must now turn to tedious evaluations
of the substitution rules \eqref{eq:mapping-flat-covariant} from
the radial gauge to the covariant representation for the sucessives
derivatives, using \eqref{eq:derivative-substitution-def} and its
higher orders.

\subsection{Substitutions for the derivatives\label{subsec:Substitutions-for-the-derivatives}}

This section establishes the substitution rules between the gauge-fixed
$O^{\left(Z\right)}\left(p,x\right)$ and the gauge-covariant $O\left(p,x\right)$
Wigner transforms. Namely, it gives, order by order how the mapping
$\left.\partial_{x}^{n}O^{\left(Z\right)}\right|_{Z=x}$ works for
$n=1$, $2$ to all order in the covariant derivative of the gauge
field and for $n=3$ up to the fourth order in gradient only. Details
are cumbersome, and can be skip at first reading.

Our first goal is to evaluate $\left.\partial_{Z}O^{\left(Z\right)}\right|_{Z=x}$
in the expression \eqref{eq:derivative-substitution-def} in order
to establish the link between the gauge-fixed derivative $\partial_{x}O^{\left(Z\right)}$
and the gauge-covariant one. The pivotal formula is \eqref{eq:DU}
which we rewrite as 
\begin{equation}
\dfrac{DU\left(b,a\right)}{\partial b^{\alpha}}U\left(a,b\right)=\mathbf{i}\mathbb{F}_{\alpha}^{\left(b\right)}\left(b,a\right)
\end{equation}
introducing the notation
\begin{equation}
\mathbb{F}_{\alpha}^{\left(c\right)}\left(b,a\right)=\left(b-a\right)^{\beta}\int_{0}^{1}ds\left[sU\left(c,\tau_{s}\right)F_{\beta\alpha}\left(\tau_{s}\right)U\left(\tau_{s},c\right)\right]\;;\;\tau_{s}=a+\left(b-a\right)s\label{eq:F-bold}
\end{equation}
for convenience. The exponant notation is fixed by the property $\mathbb{F}_{\alpha}^{\left(y\right)}\left(b,a\right)=U\left(y,x\right)\mathbb{F}_{\alpha}^{\left(x\right)}\left(b,a\right)U\left(x,y\right)$
under a gauge transformation. 

From the definition \eqref{eq:OZ} for the radial gauge Wigner transform,
one gets 
\begin{multline}
\dfrac{\partial O^{\left(Z\right)}}{\partial Z^{\alpha}}=\int d\mathfrak{x}\left[e^{-\mathbf{i}p\cdot\mathfrak{x}}\dfrac{\partial U\left(b,a\right)}{\partial b^{\alpha}}U\left(a,b\right)U\left(Z,x+\dfrac{\mathfrak{x}}{2}\right)O\left(x+\dfrac{\mathfrak{x}}{2},x-\dfrac{\mathfrak{x}}{2}\right)U\left(x-\dfrac{\mathfrak{x}}{2},Z\right)\right]_{a=x+\mathfrak{x}/2}^{b=Z}\\
+\int d\mathfrak{x}\left[e^{-\mathbf{i}p\cdot\mathfrak{x}}U\left(Z,x+\dfrac{\mathfrak{x}}{2}\right)O\left(x+\dfrac{\mathfrak{x}}{2},x-\dfrac{\mathfrak{x}}{2}\right)U\left(x-\dfrac{\mathfrak{x}}{2},Z\right)U\left(b,a\right)\dfrac{\partial U\left(a,b\right)}{\partial b^{\alpha}}\right]_{a=x-\mathfrak{x}/2}^{b=Z}
\end{multline}
using the property $U\left(a,b\right)U\left(b,a\right)=1$ of the
parallel transport. The terms written as $U\left(b,a\right)$ and
its derivatives can be extracted from the integral using the formal
property
\begin{equation}
\int d\mathfrak{x}\left[e^{-\mathbf{i}p\cdot\mathfrak{x}}f\left(\mathfrak{x}\right)U\left(Z,x+\dfrac{\mathfrak{x}}{2}\right)O\left(x+\dfrac{\mathfrak{x}}{2},x-\dfrac{\mathfrak{x}}{2}\right)U\left(x-\dfrac{\mathfrak{x}}{2},Z\right)\right]=f\left(\mathbf{i}\partial_{p}\right)O^{\left(Z\right)}\left(p,x\right)\label{eq:pseudo-Bopp-substitution}
\end{equation}
valid for any analytical function $f$. The same works for the function
$f\left(\mathfrak{x}\right)$ applied on the right, with $\partial_{p}^{\dagger}$-derivatives
instead of $\partial_{p}$. Next, the unitarity of the parallel transport
gives $\partial_{b}U\left(a,b\right)U\left(b,a\right)=-U\left(a,b\right)\partial_{b}U\left(b,a\right)$,
and so one gets
\begin{equation}
\dfrac{\partial O^{\left(Z\right)}}{\partial Z^{\alpha}}=\mathbf{i}\left[A_{\alpha}\left(Z\right)+\mathbb{F}_{\alpha}^{\left(Z\right)}\left(Z,x+\dfrac{\mathbf{i}}{2}\left(\partial_{p}-\partial_{p}^{\dagger}\right)\right),O^{\left(Z\right)}\left(p,x\right)\right]\label{eq:dOZ-dZ}
\end{equation}
where the limit $Z\rightarrow x$ can be now taken straighforwardly,
since there is no more $x$-derivatives in the right-hand side of
\eqref{eq:dOZ-dZ}. In particular, one can take the argument $\partial_{p}-\partial_{p}^{\dagger}$
in the expression \eqref{eq:dOZ-dZ} since $\mathbb{F}$ is independent
of the momentum. Relation \eqref{eq:derivative-substitution-def}
finally reads
\begin{align}
\left.\dfrac{\partial O^{\left(Z\right)}}{\partial x^{\alpha}}\right|_{Z=x} & =\dfrac{\mathfrak{D}O\left(p,x\right)}{\partial x^{\alpha}}-\mathbf{i}\left[\mathbb{F}_{\alpha}^{\left(x\right)}\left(x,x+\dfrac{\mathbf{i}}{2}\left(\partial_{p}-\partial_{p}^{\dagger}\right)\right),O\left(p,x\right)\right]\nonumber \\
 & =\dfrac{\mathbb{D}O\left(p,x\right)}{\partial x^{\alpha}}\label{eq:substitution-dOZ-DO}
\end{align}
where the right-hand side is gauge-covariant, as required. The relation
\eqref{eq:substitution-dOZ-DO} between the radial gauged and the
covariant representation of $O\left(p,x\right)$ can be injected in
the Moyal expansion to get the non-Abelian gauge-covariant gradient
expansion at first order, following the mapping \eqref{eq:mapping-flat-covariant}.
To do that one should expand $\mathbb{F}_{\alpha}^{\left(x\right)}$
in power of $\partial_{p}-\partial_{p}^{\dagger}$. Using the substitution
$U\left(x,x+z\right)F\left(x+z\right)U\left(x+z,x\right)=e^{z\cdot\mathfrak{D}_{x}}F\left(x\right)$
proved in \citep{Elze1986}, one formally has 
\begin{equation}
\mathbb{F}_{\alpha}^{\left(x\right)}\left(x,x+z\right)=\int_{0}^{1}ds\left[\left(1-s\right)e^{s\left(z\cdot\mathfrak{D}_{x}\right)}F_{\alpha\beta}\left(x\right)\right]z^{\beta}=\sum_{n=0}^{\infty}\dfrac{\left(z\cdot\mathfrak{D}_{x}\right)^{n}}{\left(n+2\right)!}F_{\alpha\beta}\left(x\right)z^{\beta}\label{eq:F-bold-series}
\end{equation}
as a covariant expansion in power of the gradient of the gauge field.
We used the formula 
\begin{equation}
\int_{0}^{1}ds\left[s^{m}e^{sx}\right]=\dfrac{\partial^{m}}{\partial x^{m}}\int_{0}^{1}ds\left[e^{sx}\right]=\sum_{n=0}^{\infty}\dfrac{x^{n}}{n!\left(n+m+1\right)}
\end{equation}
to evaluate all integrals of the type of $\mathbb{F}$ and its derivatives
(yet to come).

We thus reusse in providing an explicit method to get a completely
gauge-covariant Moyal expansion by susbtituting the usual gradient
expansion with covariant contributions according to the general recipe
\eqref{eq:mapping-flat-covariant} with
\begin{equation}
\left.\dfrac{\partial O^{\left(Z\right)}}{\partial x^{\alpha}}\right|_{Z=x}=\dfrac{\mathfrak{D}O\left(p,x\right)}{\partial x^{\alpha}}+\left[\sum_{n=0}^{\infty}\dfrac{\mathbf{i}^{n}\left(z\cdot\mathfrak{D}_{x}\right)^{n}}{2^{n+1}\left(n+2\right)!}F_{\alpha\beta}\left(x\right)z^{\beta},O\left(p,x\right)\right]_{z=\partial_{p}-\partial_{p}^{\dagger}}\label{eq:substitution-dOZ-DO-expansion}
\end{equation}
for the first order. The $\mathfrak{D}_{x}$-derivative in the commutator
applies only on the gauge field $F_{\alpha\beta}$, whereas the $\partial_{p}$-derivatives
apply only on $O\left(p,x\right)$ inside the commutator. Nevertheless,
even though the first-order derivative \eqref{eq:substitution-dOZ-DO}
can be formally writen in term of a formal covariant derivative when
$\mathbb{F}_{\alpha}^{\left(x\right)}$ is interpreted as a pseudo-gauge
potential, higher order derivatives $\left.\partial_{x}^{n}O^{\left(Z\right)}\right|_{Z=x}$
can not be straightforwardly replaced by higher derivatives $\mathbb{D}^{n}O\left(p,x\right)$.

To see that one just has to calculate the next order, namely we will
use \eqref{eq:derivative-substitution-def} to get 
\begin{multline}
\dfrac{\partial^{2}O\left(p,x\right)}{\partial x^{\alpha}\partial x^{\beta}}=\left.\left(\dfrac{\partial}{\partial x^{\alpha}}+\dfrac{\partial}{\partial Z^{\alpha}}\right)\left(\dfrac{\partial}{\partial x^{\beta}}+\dfrac{\partial}{\partial Z^{\beta}}\right)O^{\left(Z\right)}\left(p,x\right)\right|_{Z=x}=\\
=\left.\left(\dfrac{\partial}{\partial x^{\alpha}}+\dfrac{\partial}{\partial Z^{\alpha}}\right)\left(\dfrac{\partial O^{\left(Z\right)}}{\partial x^{\beta}}+\mathbf{i}\left[A_{\beta}\left(Z\right)+\mathbb{F}_{\beta}^{\left(Z\right)}\left(Z,x+\dfrac{\mathbf{i}}{2}\left(\partial_{p}-\partial_{p}^{\dagger}\right)\right),O^{\left(Z\right)}\right]\right)\right|_{Z=x}
\end{multline}
where we used \eqref{eq:dOZ-dZ} in the second line. The relation
\eqref{eq:dOZ-dZ} can be used again to evaluate $\partial_{Z}\partial_{x}O^{\left(Z\right)}=\partial_{x}\partial_{Z}O^{\left(Z\right)}$.
From the property \eqref{eq:derivative-substitution-def}, one has
\begin{multline}
\left.\left(\dfrac{\partial}{\partial x^{\alpha}}+\dfrac{\partial}{\partial Z^{\alpha}}\right)\left[A_{\beta}\left(Z\right)+\mathbb{F}_{\beta}^{\left(Z\right)}\left(Z,x+\dfrac{\mathbf{i}}{2}\left(\partial_{p}-\partial_{p}^{\dagger}\right)\right),O^{\left(Z\right)}\right]\right|_{Z=x}=\\
=\dfrac{\partial}{\partial x^{\alpha}}\left[A_{\beta}\left(x\right)+\mathbb{F}_{\alpha}^{\left(x\right)}\left(x,x+\dfrac{\mathbf{i}}{2}\left(\partial_{p}-\partial_{p}^{\dagger}\right)\right),O\left(p,x\right)\right]
\end{multline}
and so 
\begin{multline}
\dfrac{\partial}{\partial x^{\alpha}}\left(O\left(p,x\right)-\mathbf{i}\left[A_{\beta}\left(x\right)+\mathbb{F}_{\alpha}^{\left(x\right)}\left(x,x+\dfrac{\mathbf{i}}{2}\left(\partial_{p}-\partial_{p}^{\dagger}\right)\right),O\left(p,x\right)\right]\right)=\\
=\left.\left(\dfrac{\partial}{\partial x^{\alpha}}+\dfrac{\partial}{\partial Z^{\alpha}}\right)\dfrac{\partial O^{\left(Z\right)}}{\partial x^{\beta}}\right|_{Z=x}=\\
=\left.\dfrac{\partial^{2}O^{\left(Z\right)}}{\partial x^{\alpha}\partial x^{\beta}}\right|_{Z=x}+\mathbf{i}\left.\dfrac{\partial}{\partial x^{\beta}}\left[A_{\alpha}\left(Z\right)+\mathbb{F}_{\alpha}^{\left(Z\right)}\left(Z,x+\dfrac{\mathbf{i}}{2}\left(\partial_{p}-\partial_{p}^{\dagger}\right)\right),O^{\left(Z\right)}\left(p,x\right)\right]\right|_{Z=x}
\end{multline}
now expanding the derivative $\partial_{\beta}$ on the right-hand
side, and because $\partial_{x}A\left(Z\right)=0$, and finally using
\eqref{eq:substitution-dOZ-DO} to evaluate $\left.\partial_{\beta}O^{\left(Z\right)}\right|_{Z=x}$,
one gets the substitution rule for the second order derivative
\begin{equation}
\left.\dfrac{\partial^{2}O^{\left(Z\right)}}{\partial x^{\alpha}\partial x^{\beta}}\right|_{Z=x}=\dfrac{\mathbb{D}^{2}O\left(p,x\right)}{\partial x^{\alpha}\partial x^{\beta}}-\mathbf{i}\left[\dfrac{\partial}{\partial x^{\beta}}\mathbb{F}_{\alpha}^{\left(Z\right)}\left(Z,x+z\right),O^{\left(Z\right)}\left(p,x\right)\right]_{Z=x}^{z=\mathbf{i}\left(\partial_{p}-\partial_{p}^{\dagger}\right)/2}\label{eq:substitution-ddOZ-DDO}
\end{equation}
which thus misses to be a simple replacement of the $\left.\partial_{x}^{n}O^{\left(Z\right)}\right|_{Z=x}$
by the formal covariant derivative $\mathbb{D}_{x}^{n}O\left(p,x\right)$.
However, a rapid calculation gives 
\begin{multline}
\left.\dfrac{\partial}{\partial x^{\beta}}\mathbb{F}_{\alpha}^{\left(Z\right)}\left(Z,x+z\right)\right|_{Z=x}=\\
=\int_{0}^{1}ds\left[s\left(1-s\right)U\left(b,a\right)\left(\dfrac{\mathfrak{D}F_{\alpha\gamma}\left(a\right)}{\partial a^{\beta}}-\mathbf{i}\left[\mathbb{F}_{\beta}^{\left(a\right)}\left(a,b\right),F_{\alpha\gamma}\left(a\right)\right]\right)U\left(a,b\right)\right]_{b=x}^{a=x+zs}z^{\gamma}\label{eq:d-F-bold}
\end{multline}
and thus shows that all terms in \eqref{eq:substitution-ddOZ-DDO}
are gauge covariant as required. A term proportionnal to $F_{\alpha\beta}$
has been discarded in \eqref{eq:d-F-bold}, since it will not appear
in the Moyal expansion with contracted indices, i.e. the interchange
$\alpha\leftrightarrow\beta$ is allowed at any stage of the computation
to simplify it. Though we just missed to give a formal resummation
of all the covariant contributions appearing in the Moyal expansion,
we clearly succeed in establishing a gauge covariant method to get
the Moyal expansion order by order using the mapping \eqref{eq:mapping-flat-covariant}. 

Using the same method we used to get \eqref{eq:F-bold-series}, one
can evaluate 
\begin{equation}
\mathbb{F}_{\beta}^{\left(x+zs\right)}\left(x+zs,z\right)=U\left(x+zs,x\right)\left[\sum_{n=0}^{\infty}\dfrac{n+1}{\left(n+2\right)!}\left(sz\cdot\mathfrak{D}_{x}\right)^{n}F_{\delta\beta}\left(x\right)z^{\delta}\right]U\left(x+zs,x\right)
\end{equation}
using $U\left(a,b\right)U\left(b,c\right)=U\left(a,c\right)$ to evaluate
the term into brackets. The above contribution nevertheless does not
contribute to the covariant Moyal expansion, because both pairs of
indices $\left(\delta,\gamma\right)$ and $\left(\alpha,\beta\right)$
are dummy indices in the Moyal expansion, whereas there will be some
commutators of the form $\left[F_{\alpha\gamma},F_{\beta\delta}\right]$
and their covariant derivatives in \eqref{eq:d-F-bold}. Thus \eqref{eq:d-F-bold}
reads
\begin{equation}
\left.\dfrac{\partial}{\partial x^{\beta}}\mathbb{F}_{\alpha}^{\left(Z\right)}\left(Z,x+z\right)\right|_{Z=x}=\sum_{n=0}^{\infty}\dfrac{n+1}{\left(n+3\right)!}\left(z\cdot\mathfrak{D}_{x}\right)^{n}\dfrac{\mathfrak{D}F_{\alpha\gamma}\left(x\right)}{\partial x^{\beta}}z^{\gamma}+\cdots\label{eq:d-F-bold-expansion}
\end{equation}
being understood that the covariant derivatives $\mathfrak{D}_{x}$
apply only on the gauge-field next to them, and the forgoten terms
in $\cdots$ do not participate in the Moyal expansion. The expression
\eqref{eq:substitution-ddOZ-DDO} can thus be rewriten as 
\begin{multline}
\left.\dfrac{\partial^{2}O^{\left(Z\right)}}{\partial x^{\alpha}\partial x^{\beta}}\right|_{Z=x}\overset{\text{sym.}}{=}\dfrac{\mathfrak{D}^{2}O\left(p,x\right)}{\partial x^{\alpha}\partial x^{\beta}}+\sum_{n=0}^{\infty}\left[\left(\dfrac{\mathbf{i}}{2}\right)^{n}\dfrac{n+2}{\left(n+3\right)!}\left(z^{\delta}\dfrac{\mathfrak{D}}{\partial x^{\delta}}\right)^{n}\dfrac{\mathfrak{D}F_{\beta\gamma}\left(x\right)}{\partial x^{\alpha}}z^{\gamma},O\left(p,x\right)\right]_{z=\partial_{p}-\partial_{p}^{\dagger}}\\
+\sum_{n=0}^{\infty}\left[\dfrac{\mathbf{i}^{n}}{2^{n+1}\left(n+2\right)!}\left(z^{\delta}\dfrac{\mathfrak{D}}{\partial x^{\delta}}\right)^{n}F_{\beta\gamma}\left(x\right)z^{\gamma},\dfrac{\mathfrak{D}O\left(p,x\right)}{\partial x^{\alpha}}\right]_{z=\partial_{p}-\partial_{p}^{\dagger}}+\\
+\sum_{n=0}^{\infty}\sum_{m=0}^{\infty}\dfrac{\left(\mathbf{i}/2\right)^{n+m}}{4\left(n+2\right)!\left(m+2\right)!}\left[\left(z^{\epsilon}\dfrac{\mathfrak{D}}{\partial x^{\epsilon}}\right)^{n}F_{\beta\delta}z^{\delta},\left[\left(z^{\varepsilon}\dfrac{\mathfrak{D}}{\partial x^{\varepsilon}}\right)^{m}F_{\alpha\gamma}z^{\gamma},O\right]\right]_{z=\partial_{p}-\partial_{p}^{\dagger}}\label{eq:substitution-ddOZ-DDO-expansion}
\end{multline}
valid at all order in the covariant derivatives of the gauge fields,
once injected in the covariant Moyal expansion following the recipe
\eqref{eq:mapping-flat-covariant}. One more time, let me repeat that
the above expansion neglects terms odd in the permutation $\alpha\leftrightarrow\beta$
and/or $\gamma\leftrightarrow\delta$, since these contributions would
vanish in the Moyal expansion. This is what I call the 'symmetrized
equality' $\overset{\text{sym.}}{=}$. There was no term forgoten
in the first order calulation, and expression \eqref{eq:substitution-dOZ-DO-expansion}
was exact.

The above procedure can be performed up to any order. Higher order
terms are nevertheless more and more cumbersome. Unfortunately, to
truncate the expansion at order four in gradient, we need the next
order term as well. To understand why, have a look at \eqref{eq:d-F-bold-expansion}.
Especially the term of the form $\mathfrak{D}F\cdot z$, which gives
a contribution $\mathfrak{D}F\cdot\partial_{p}O$ in the substitution
\eqref{eq:substitution-ddOZ-DDO-expansion} for the second order derivative
$\left.\partial_{x}^{2}O^{\left(Z\right)}\right|_{Z=x}$. In fact,
at any order in the evaluation of $\left.\partial_{x}^{n}O^{\left(Z\right)}\right|_{Z=x}$,
a term of the form $\mathfrak{D}^{n-1}F\cdot\partial_{p}O$ will be
generated. This is clearly of order $n$ in gradient, provided we
count the gradient of the gauge field as well. In general, one prefers
to count the powers of the derivatives of the operator $O$, and so
to take the correct contribution at order four in the gradients of
$O$, one must include a contribution $\mathfrak{D}^{2}F\cdot\partial_{p}O$
from the third order derivative $\left.\partial_{x}^{3}O^{\left(Z\right)}\right|_{Z=x}$.

This will be the unique contribution we need to evaluate. Indeed,
using schematic representation, one has, from the substitution rules
from the Moyal expansion in the radial gauge to its gauge covariant
representation
\begin{equation}
\partial_{x}O_{1}^{\left(Z\right)}\cdot\partial_{p}O_{2}^{\left(Z\right)}\rightarrow\partial_{x}O_{1}\cdot\partial_{p}O_{2}+F\cdot\partial_{p}O_{1}\cdot\partial_{p}O_{2}+\mathfrak{D}F\cdot\partial_{p}^{2}O_{1}\cdot\partial_{p}O_{2}+\cdots
\end{equation}
up to higher order in gradients. This term is of first order in the
$\star_{0}$-expansion, but second order in gradient, since we count
every gradient applied on either $O_{1}$ or $O_{2}$ in the later
case. The term $\partial_{p}O_{1}^{\left(Z\right)}\cdot\partial_{x}O_{2}^{\left(Z\right)}$
looks schematically the same, and so we disregard it. At the second
order in the $\star_{0}$-expansion, one has terms of the form of
fourth order gradient, e.g. 
\begin{equation}
\partial_{x}^{2}O_{1}^{\left(Z\right)}\cdot\partial_{p}^{2}O_{2}^{\left(Z\right)}\rightarrow\mathfrak{D}F\cdot\partial_{p}O_{1}\cdot\partial_{p}^{2}O_{2}+\mathfrak{D}^{2}F\cdot\partial_{p}^{2}O_{1}\cdot\partial_{p}^{2}O_{2}+\mathfrak{D}^{2}O_{1}\cdot\partial_{p}^{2}O_{2}+\cdots
\end{equation}
and terms of the form 
\begin{equation}
\partial_{x}\partial_{p}O_{1}^{\left(Z\right)}\cdot\partial_{x}\partial_{p}O_{2}^{\left(Z\right)}\rightarrow\mathfrak{D}\partial_{p}O_{1}\cdot\mathfrak{D}\partial_{p}O_{2}+F\cdot\partial_{p}^{2}O_{1}\cdot\mathfrak{D}\partial_{p}O_{2}+\cdots
\end{equation}
so there is a term $\mathfrak{D}F\cdot\partial_{p}O_{1}\cdot\partial_{p}^{2}O_{2}$
of the third order in gradient appearing when taking the mapping \eqref{eq:mapping-flat-covariant}
of the second order of the $\star_{0}$-expansion (and fourth order
in gradient). This is an anomalous term which avoids to resum all
the expansion in a convenient way using the method presented here.

At the next order (third in $\star_{0}$, and sixth in gradient),
one has terms of the form
\begin{equation}
\partial_{x}^{3}O_{1}^{\left(Z\right)}\cdot\partial_{p}^{3}O_{2}^{\left(Z\right)}\rightarrow\mathfrak{D}^{2}F\cdot\partial_{p}O_{1}\cdot\partial_{p}^{3}O_{2}+\mathfrak{D}^{3}F\cdot\partial_{p}^{2}O_{1}\cdot\partial_{p}^{3}O_{2}+\cdots+\mathfrak{D}^{3}O_{1}\cdot\mathfrak{D}^{3}O_{2}+\cdots
\end{equation}
where a fourth order term in gradient thus appear. The other contributions
in the $\star_{0}$ expansion, e.g.
\begin{equation}
\partial_{x}^{2}\partial_{p}O_{1}^{\left(Z\right)}\cdot\partial_{x}\partial_{p}^{2}O_{2}^{\left(Z\right)}\rightarrow\mathfrak{D}F\cdot\partial_{p}^{2}O_{1}\cdot\mathfrak{D}\partial_{p}^{2}O_{1}+\cdots
\end{equation}
will be of higher power in gradient (here we neglect terms of order
5 in gradient).

Thus, the price to pay for the use of the mapping \eqref{eq:mapping-flat-covariant}
allowing the transformation from the flat Moyal product $\star_{0}$
to the covariant one $\star$ is a cumbersome counting of the order
in a gradient expansion. Since we want to stop our calculation to
the fourth order in gradients, we will not give the full calculation
of the evaluation of the third order in the $\star_{0}$-expansion
\begin{equation}
\star_{0}=1+\dfrac{\mathbf{i}}{2}\left(\partial_{x}^{\dagger}\partial_{p}-\partial_{p}^{\dagger}\partial_{x}\right)+\dfrac{1}{2}\left(\dfrac{\mathbf{i}}{2}\right)^{2}\left(\partial_{x}^{\dagger}\partial_{p}-\partial_{p}^{\dagger}\partial_{x}\right)^{2}+\dfrac{1}{6}\left(\dfrac{\mathbf{i}}{2}\right)^{3}\left(\partial_{x}^{\dagger}\partial_{p}-\partial_{p}^{\dagger}\partial_{x}\right)^{3}+\cdots
\end{equation}
namely of the term $\left.\partial_{x}^{3}O^{\left(Z\right)}\right|_{Z=x}$
but give the term of the form $\mathfrak{D}^{2}F\cdot\partial_{p}O$
only. The third order derivative looks like
\begin{multline}
\dfrac{\partial^{3}O\left(p,x\right)}{\partial x^{\alpha}\partial x^{\beta}\partial x^{\gamma}}=\left.\left(\dfrac{\partial}{\partial x^{\alpha}}+\dfrac{\partial}{\partial Z^{\alpha}}\right)\left(\dfrac{\partial}{\partial x^{\beta}}+\dfrac{\partial}{\partial Z^{\beta}}\right)\left(\dfrac{\partial}{\partial x^{\gamma}}+\dfrac{\partial}{\partial Z^{\gamma}}\right)O^{\left(Z\right)}\left(p,x\right)\right|_{Z=x}\\
=\left.\left(\dfrac{\partial}{\partial x^{\alpha}}+\dfrac{\partial}{\partial Z^{\alpha}}\right)\left(\dfrac{\partial}{\partial x^{\beta}}+\dfrac{\partial}{\partial Z^{\beta}}\right)\left(\dfrac{\partial O^{\left(Z\right)}}{\partial x^{\gamma}}+\mathbf{i}\left[\mathbb{F}_{\gamma}^{\left(Z\right)}\left(Z,x+z\right),O^{\left(Z\right)}\right]\right)\right|_{Z=x}^{z=\mathbf{i}\left(\partial_{p}-\partial_{p}^{\dagger}\right)/2}+\mathcal{O}\left(\partial^{2}\right)
\end{multline}
where we can neglect the terms in $A_{\alpha}$ since they will be
of higher power in gradient (they would generate some covariant contribution
$\mathfrak{D}^{3}O$ when they are all taken into account properly
since the expansion is covariant -- we disregard these terms as being
of higher order in gradient), as well as the terms $\partial_{Z}\mathbb{F}$
since they do not generate any term of the form $\mathfrak{D}F$ and
will thus be of higher power in $\partial_{p}$ as well. Now, we use
the definition \eqref{eq:dOZ-dZ} to substitute $\partial_{Z}O^{\left(Z\right)}$
with some $\left[\mathbb{F}^{\left(Z\right)},O^{\left(Z\right)}\right]$,
and we obtain 
\begin{equation}
\left.\dfrac{\partial^{3}O^{\left(Z\right)}\left(p,x\right)}{\partial x^{\alpha}\partial x^{\beta}\partial x^{\gamma}}\right|_{Z=x}=\left.-3\mathbf{i}\left[\dfrac{\partial^{2}}{\partial x^{\alpha}\partial x^{\beta}}\mathbb{F}_{\gamma}^{\left(Z\right)}\left(Z,x+z\right),O^{\left(Z\right)}\right]\right|_{Z=x}^{z=\mathbf{i}\left(\partial_{p}-\partial_{p}^{\dagger}\right)/2}+\mathcal{O}\left(\partial^{2}\right)
\end{equation}
at the leading order in gradient. We used that the Moyal expansion
at third order gives a contribution $\partial_{x}^{3}O_{1}\partial_{p}^{3}O_{2}$
and therefore the indices $\alpha$, $\beta$ and $\gamma$ are all
contracted in the final result. There are thus three terms of the
form $\partial_{x}^{2}\mathbb{F}$ since we discuss the third order
derivatives. There were in fact two terms of the form \eqref{eq:d-F-bold}
in the second order substitution \eqref{eq:substitution-ddOZ-DDO},
one being hidden in the expression of $\mathbb{D}^{2}O\left(p,x\right)$
as one can check by expansion of the second order of \eqref{eq:substitution-dOZ-DO}. 

Evaluating the second derivative of $\mathbb{F}$ before taking the
limit $Z\rightarrow x$ gives, at the leading order 
\begin{multline}
\left.\dfrac{\partial^{3}O^{\left(Z\right)}\left(p,x\right)}{\partial x^{\alpha}\partial x^{\beta}\partial x^{\gamma}}\right|_{Z=x}=\dfrac{3}{2}\int_{0}^{1}ds\left[s\left(1-s\right)^{2}\right]\left[\dfrac{\mathfrak{D}^{2}F_{\gamma\delta}\left(x\right)}{\partial x^{\alpha}\partial x^{\beta}}\left(\partial_{p}-\partial_{p}^{\dagger}\right)^{\delta},O\left(p,x\right)\right]+\mathcal{O}\left(\partial^{2}\right)\\
=\dfrac{1}{8}\left\{ \dfrac{\mathfrak{D}^{2}F_{\gamma\delta}\left(x\right)}{\partial x^{\alpha}\partial x^{\beta}},\dfrac{\partial O\left(p,x\right)}{\partial p_{\delta}}\right\} +\mathcal{O}\left(\partial^{2}\right)\label{eq:substitution-dddOZ-DDDO}
\end{multline}
which should be injected in the gradient expansion when one wants
to retain fourth order terms in gradient.

We are now ready to give the expansion of the non-Abelian gauge covariant
Moyal product at fourth order in gradient. 

\subsection{Gauge-covariant gradient expansion\label{subsec:Gauge-covariant-gradient-expansion}}

To get the gradient expansion at the desired order, we simply expand
the different contributions \eqref{eq:substitution-dOZ-DO} and \eqref{eq:substitution-ddOZ-DDO}
up to the fourth order in gradient, and add the anomalous contribution
\eqref{eq:substitution-dddOZ-DDDO} coming from the higher order. 

The expansion of the mapping \eqref{eq:mapping-flat-covariant} at
first order in derivatives is already explicitly given in \eqref{eq:substitution-dOZ-DO-expansion}.
One has 
\begin{multline}
\left.\dfrac{\partial O^{\left(Z\right)}\left(p,x\right)}{\partial x^{\alpha}}\right|_{Z=x}=\dfrac{\mathfrak{D}O\left(p,x\right)}{\partial x^{\alpha}}+\dfrac{1}{4}\left\{ F_{\alpha\delta}\left(x\right),\dfrac{\partial O}{\partial p_{\delta}}\right\} +\\
+\dfrac{\mathbf{i}}{24}\left[\dfrac{\mathfrak{D}F_{\alpha\delta}}{\partial x^{\gamma}},\dfrac{\partial^{2}O\left(p,x\right)}{\partial p_{\gamma}\partial p_{\delta}}\right]-\dfrac{1}{192}\left\{ \dfrac{\mathfrak{D}^{2}F_{\alpha\delta}}{\partial x^{\beta}\partial x^{\gamma}},\dfrac{\partial^{3}O\left(p,x\right)}{\partial p_{\beta}\partial p_{\gamma}\partial p_{\delta}}\right\} +\mathcal{O}\left(\partial^{4}\right)\label{eq:substitution-dxOZ-1}
\end{multline}
where the alternance of anti-commutators and commutators comes from
the structure of \eqref{eq:substitution-dOZ-DO-expansion} in power
of $\left(\partial_{p}-\partial_{p}^{\dagger}\right)$. We need the
third order in gradient, since in the expansion of $\star_{0}$, the
term $\left.\partial_{x}O^{\left(Z\right)}\right|_{Z=x}$ appears
in combination with a first order derivative $\partial_{p}O\left(p,x\right)$
which is not affected by the mapping.

For the second order derivative, one uses \eqref{eq:substitution-ddOZ-DDO-expansion}
to get 
\begin{multline}
\left.\dfrac{\partial^{2}O^{\left(Z\right)}\left(p,x\right)}{\partial x^{\alpha}\partial x^{\beta}}\right|_{Z=x}=\dfrac{\mathfrak{D}^{2}O\left(p,x\right)}{\partial x^{\alpha}\partial x^{\beta}}+\dfrac{1}{4}\left\{ F_{\beta\gamma}\left(x\right),\dfrac{\mathfrak{D}\partial O\left(p,x\right)}{\partial x^{\alpha}\partial p_{\gamma}}\right\} +\dfrac{1}{3}\left\{ \dfrac{\mathfrak{D}F_{\beta\gamma}}{\partial x^{\alpha}},\dfrac{\partial O}{\partial p_{\gamma}}\right\} \\
+\dfrac{\mathbf{i}}{16}\left[\dfrac{\mathfrak{D}^{2}F_{\beta\gamma}}{\partial x^{\alpha}\partial x^{\delta}},\dfrac{\partial^{2}O\left(p,x\right)}{\partial p_{\gamma}\partial p_{\delta}}\right]+\dfrac{1}{16}\left\{ F_{\alpha\delta}\left(x\right),\left\{ F_{\beta\gamma}\left(x\right),\dfrac{\partial^{2}O\left(p,x\right)}{\partial p_{\gamma}\partial p_{\delta}}\right\} \right\} +\mathcal{O}\left(\partial^{3}\right)\label{eq:substitution-dxOZ-2}
\end{multline}
with the anomalous term $\mathfrak{D}F\cdot\partial_{p}O$ discussed
in section \ref{subsec:Substitutions-for-the-derivatives}. Note that
the covariant derivative $\mathfrak{D}_{x}$ and the momentum derivative
$\partial_{p}$ commute. 

Finally, the anomalous term at the third order \eqref{eq:substitution-dddOZ-DDDO}
should be taken in order to get the complete gauge-covariant Moyal
expansion up to the fourth order. The covariant Moyal product then
reads
\begin{multline}
O_{1}\left(p,x\right)\star O_{2}\left(p,x\right)=O_{1}\left(p,x\right)O_{2}\left(p,x\right)+\dfrac{\mathbf{i}}{2}\left(\left.\dfrac{\partial O_{1}^{\left(Z\right)}}{\partial x^{\alpha}}\right|_{Z=x}\dfrac{\partial O_{2}}{\partial p_{\alpha}}-\dfrac{\partial O_{1}}{\partial p_{\alpha}}\left.\dfrac{\partial O_{2}^{\left(Z\right)}}{\partial x^{\alpha}}\right|_{Z=x}\right)+\\
+\dfrac{1}{2}\left(\dfrac{\mathbf{i}}{2}\right)^{2}\left(\left.\dfrac{\partial^{2}O_{1}^{\left(Z\right)}}{\partial x^{\alpha}\partial x^{\beta}}\right|_{Z=x}\dfrac{\partial^{2}O_{2}}{\partial p_{\alpha}\partial p_{\beta}}-2\dfrac{\partial}{\partial p_{\alpha}}\left.\dfrac{\partial O_{1}^{\left(Z\right)}}{\partial x^{\beta}}\right|_{Z=x}\dfrac{\partial}{\partial p_{\beta}}\left.\dfrac{\partial O_{2}^{\left(Z\right)}}{\partial x^{\alpha}}\right|_{Z=x}+\dfrac{\partial^{2}O_{1}}{\partial p_{\alpha}\partial p_{\beta}}\left.\dfrac{\partial^{2}O_{2}^{\left(Z\right)}}{\partial x^{\alpha}\partial x^{\beta}}\right|_{Z=x}\right)\\
+\dfrac{1}{6}\left(\dfrac{\mathbf{i}}{2}\right)^{3}\left(\left.\dfrac{\partial^{3}O_{1}^{\left(Z\right)}}{\partial x^{\alpha}\partial x^{\beta}\partial x^{\gamma}}\right|_{Z=x}\dfrac{\partial^{3}O_{2}}{\partial p_{\alpha}\partial p_{\beta}\partial p_{\gamma}}-\dfrac{\partial^{3}O_{1}}{\partial p_{\alpha}\partial p_{\beta}\partial p_{\gamma}}\left.\dfrac{\partial^{3}O_{2}^{\left(Z\right)}}{\partial x^{\alpha}\partial x^{\beta}\partial x^{\gamma}}\right|_{Z=x}\right)+\mathcal{O}\left(\partial^{5}\right)
\end{multline}
where the other terms of the expansion $\left(\partial_{x}^{\dagger}\partial_{p}-\partial_{p}^{\dagger}\partial_{x}\right)^{3}$
do not contribute to the four order in gradient, as explained after
\eqref{eq:substitution-ddOZ-DDO-expansion}. We just have to substitute
\eqref{eq:substitution-dxOZ-1}, \eqref{eq:substitution-dxOZ-2} and
\eqref{eq:substitution-dddOZ-DDDO} in the above expansion to get
the covariant Moyal product.

After a few algebra, and classifying the terms order by order in gradients,
one finally has 
\begin{equation}
O_{1}\left(p,x\right)\star O_{2}\left(p,x\right)=O_{1}\left(p,x\right)O_{2}\left(p,x\right)+\mathcal{O}_{\star}\left(\partial^{2}\right)+\mathcal{O}_{\star}\left(\partial^{3}\right)+\mathcal{O}_{\star}\left(\partial^{4}\right)+\mathcal{O}\left(\partial^{5}\right)\label{eq:STAR-PRODUCT}
\end{equation}
where the first order term 
\begin{multline}
\mathcal{O}_{\star}\left(\partial^{2}\right)=\dfrac{\mathbf{i}}{2}\left(\dfrac{\mathfrak{D}O_{1}}{\partial x^{\alpha}}\dfrac{\partial O_{2}}{\partial p_{\alpha}}-\dfrac{\partial O_{1}}{\partial p_{\alpha}}\dfrac{\mathfrak{D}O_{2}}{\partial x^{\alpha}}\right)-\\
-\dfrac{\mathbf{i}}{8}\left(F_{\alpha\beta}\left(x\right)\dfrac{\partial O_{1}}{\partial p_{\alpha}}\dfrac{\partial O_{2}}{\partial p_{\beta}}+2\dfrac{\partial O_{1}}{\partial p_{\alpha}}F_{\alpha\beta}\left(x\right)\dfrac{\partial O_{2}}{\partial p_{\beta}}+\dfrac{\partial O_{1}}{\partial p_{\alpha}}\dfrac{\partial O_{2}}{\partial p_{\beta}}F_{\alpha\beta}\left(x\right)\right)
\end{multline}
and the second order term
\begin{multline}
\mathcal{O}_{\star}\left(\partial^{3}\right)=\dfrac{1}{12}\left(\dfrac{\partial O_{1}}{\partial p_{\gamma}}\dfrac{\mathfrak{D}F_{\gamma\beta}}{\partial x^{\alpha}}\dfrac{\partial^{2}O_{2}}{\partial p_{\alpha}\partial p_{\beta}}+\dfrac{\partial^{2}O_{1}}{\partial p_{\alpha}\partial p_{\beta}}\dfrac{\mathfrak{D}F_{\gamma\beta}}{\partial x^{\alpha}}\dfrac{\partial O_{2}}{\partial p_{\gamma}}\right)+\\
+\dfrac{1}{24}\left[\dfrac{\mathfrak{D}F_{\gamma\beta}}{\partial x^{\alpha}},\dfrac{\partial^{2}O_{1}}{\partial p_{\alpha}\partial p_{\beta}}\dfrac{\partial O_{2}}{\partial p_{\gamma}}-\dfrac{\partial O_{1}}{\partial p_{\gamma}}\dfrac{\partial^{2}O_{2}}{\partial p_{\alpha}\partial p_{\beta}}\right]\label{eq:O-3}
\end{multline}
were already writen in \eqref{eq:MOYAL-EXPANSION-GAUGE}. We here
add the fourth order term
\begin{multline}
\mathcal{O}_{\star}\left(\partial^{4}\right)=-\dfrac{1}{8}O_{1}\left(p,x\right)\left(\mathfrak{D}_{x}^{\dagger}\partial_{p}-\partial_{p}^{\dagger}\mathfrak{D}_{x}\right)^{2}O_{2}\left(p,x\right)+\\
+\dfrac{1}{32}\left(\left\{ F_{\gamma\beta},\dfrac{\partial\mathfrak{D}O_{1}}{\partial x^{\alpha}\partial p_{\gamma}}\right\} \dfrac{\partial^{2}O_{2}}{\partial p_{\alpha}\partial p_{\beta}}+2\dfrac{\partial\mathfrak{D}O_{1}}{\partial x^{\alpha}\partial p_{\gamma}}\left\{ F_{\gamma\beta},\dfrac{\partial^{2}O_{2}}{\partial p_{\alpha}\partial p_{\beta}}\right\} \right)+\\
+\dfrac{1}{32}\left(2\left\{ F_{\gamma\beta},\dfrac{\partial^{2}O_{1}}{\partial p_{\alpha}\partial p_{\beta}}\right\} \dfrac{\partial\mathfrak{D}O_{2}}{\partial x^{\alpha}\partial p_{\gamma}}+\dfrac{\partial^{2}O_{1}}{\partial p_{\alpha}\partial p_{\beta}}\left\{ F_{\gamma\beta},\dfrac{\partial\mathfrak{D}O_{2}}{\partial x^{\alpha}\partial p_{\gamma}}\right\} \right)+\\
+\dfrac{\mathbf{i}}{128}\left(\left[\dfrac{\mathfrak{D}^{2}F_{\gamma\beta}}{\partial x^{\alpha}\partial x^{\delta}},\dfrac{\partial^{2}O_{1}}{\partial p_{\gamma}\partial p_{\delta}}\right]\dfrac{\partial^{2}O_{2}}{\partial p_{\alpha}\partial p_{\beta}}+\dfrac{\partial^{2}O_{1}}{\partial p_{\alpha}\partial p_{\beta}}\left[\dfrac{\mathfrak{D}^{2}F_{\gamma\beta}}{\partial x^{\alpha}\partial x^{\delta}},\dfrac{\partial^{2}O_{2}}{\partial p_{\gamma}\partial p_{\delta}}\right]\right)+\\
+\dfrac{\mathbf{i}}{3.2^{7}}\left(\dfrac{\partial^{3}O_{1}}{\partial p_{\alpha}\partial p_{\beta}\partial p_{\gamma}}\left\{ \dfrac{\mathfrak{D}^{2}F_{\gamma\delta}}{\partial x^{\alpha}\partial x^{\beta}},\dfrac{\partial O_{2}}{\partial p_{\delta}}\right\} +\left\{ \dfrac{\mathfrak{D}^{2}F_{\gamma\delta}}{\partial x^{\alpha}\partial x^{\beta}},\dfrac{\partial^{3}O_{1}}{\partial p_{\alpha}\partial p_{\beta}\partial p_{\gamma}}\right\} \dfrac{\partial O_{2}}{\partial p_{\delta}}-\left(\partial_{p}^{3}\leftrightarrow\partial_{p}\right)\right)+\\
-\dfrac{1}{128}\left(\left\{ F_{\alpha\delta},\left\{ F_{\beta\gamma},\dfrac{\partial^{2}O_{1}}{\partial p_{\gamma}\partial p_{\delta}}\right\} \right\} \dfrac{\partial^{2}O_{2}}{\partial p_{\alpha}\partial p_{\beta}}+\dfrac{\partial^{2}O_{1}}{\partial p_{\gamma}\partial p_{\delta}}\left\{ F_{\beta\gamma},\left\{ F_{\alpha\delta},\dfrac{\partial^{2}O_{2}}{\partial p_{\alpha}\partial p_{\beta}}\right\} \right\} \right)-\\
-\dfrac{1}{64}\left\{ F_{\beta\gamma},\dfrac{\partial^{2}O_{1}}{\partial p_{\gamma}\partial p_{\delta}}\right\} \left\{ F_{\alpha\delta},\dfrac{\partial^{2}O_{2}}{\partial p_{\alpha}\partial p_{\beta}}\right\} \label{eq:O-4}
\end{multline}
for the gauge-covariant Moyal expansion up to the four order in gradient
of the operator. We define $O\mathfrak{D}_{x}^{\dagger}=\partial_{x}O+\mathbf{i}\left[O,A\right]=\mathfrak{D}_{x}O$.
The symbol $\left(\partial_{p}^{3}\leftrightarrow\partial_{p}\right)$
signifies that one must permute the triple and single derivatives,
including their indices, without changing the indices of the covariant
derivative of the gauge field, see \eqref{eq:O-3} which can be noted
\begin{equation}
\mathcal{O}_{\star}\left(\partial^{3}\right)=\dfrac{1}{12}\left(\dfrac{\partial O_{1}}{\partial p_{\gamma}}\dfrac{\mathfrak{D}F_{\gamma\beta}}{\partial x^{\alpha}}\dfrac{\partial^{2}O_{2}}{\partial p_{\alpha}\partial p_{\beta}}+\left(\partial_{p}^{2}\leftrightarrow\partial_{p}\right)\right)+\dfrac{1}{24}\left[\dfrac{\mathfrak{D}F_{\gamma\beta}}{\partial x^{\alpha}},\dfrac{\partial^{2}O_{1}}{\partial p_{\alpha}\partial p_{\beta}}\dfrac{\partial O_{2}}{\partial p_{\gamma}}-\left(\partial_{p}^{2}\leftrightarrow\partial_{p}\right)\right]
\end{equation}
as well.

Expressions \eqref{eq:STAR-PRODUCT}-\eqref{eq:O-4} constitute the
first important result of this paper. It validates the method explained
in section \ref{subsec:Two-different-Wigner-transforms} aiming at
calculating a gauge-covariant Moyal product at any order.

\subsection{Abelian and pure gauge cases\label{subsec:Abelian-and-pure-gauge}}

In this section, we review some easier situations than the non-Abelian
covariant case. 

We first discuss the case of a pure gauge, characterised by a vanishing
gauge field $F_{\alpha\beta}=0$ but still having non trivial gauge
potential. This case exists only in the non-Abelian situation. 

In this case one has 
\begin{equation}
\dfrac{DU\left(b,a\right)}{\partial b^{\alpha}}U\left(a,b\right)=0\label{eq:DU-pure-gauge}
\end{equation}
and so $\mathbb{F}=0$ in all the calculations done in section \ref{subsec:Substitutions-for-the-derivatives}.
From the expansion of the $\star$-product in \eqref{eq:STAR-PRODUCT},
one might expect that 

\begin{equation}
\lim_{F\rightarrow0}\star=\exp\left[\dfrac{\mathbf{i}\hbar}{2}\left(\dfrac{\mathfrak{D}^{\dagger}}{\partial x^{\alpha}}\dfrac{\partial}{\partial p_{\alpha}}-\dfrac{\partial^{\dagger}}{\partial p_{\alpha}}\dfrac{\mathfrak{D}}{\partial x^{\alpha}}\right)\right]\label{eq:MOYAL-pure-gauge}
\end{equation}
with $O\left(p,x\right)\mathfrak{D}_{\alpha}^{\dagger}=O\left(p,x\right)\left[\partial_{\alpha}^{\dagger}+\mathbf{i}\left[\cdot,A_{\alpha}\right]\right]=\partial_{\alpha}O-\mathbf{i}\left[A_{\alpha},O\right]$.
To show that, we simply write, from the definition \eqref{eq:OZ}
of $O^{\left(Z\right)}$ 
\begin{equation}
O^{\left(Z\right)}\left(p,x\right)=U\left(Z,x\right)O\left(p,x\right)U\left(x,Z\right)\Rightarrow\dfrac{\partial O^{\left(Z\right)}}{\partial x^{\alpha}}=U\left(Z,x\right)\dfrac{\mathfrak{D}O\left(p,x\right)}{\partial x^{\alpha}}U\left(Z,x\right)
\end{equation}
since $\partial_{\alpha}U\left(x,Z\right)U\left(Z,x\right)=\mathbf{i}A_{\alpha}\left(x\right)$
in the limit of vanishing gauge field, see \eqref{eq:DU-pure-gauge}.
There is thus no more $Z$-dependency in the covariant derivative
$\mathfrak{D}O\left(p,x\right)$. We thus have generically 
\begin{equation}
\dfrac{\partial^{n}O^{\left(Z\right)}}{\partial x^{n}}=U\left(Z,x\right)\dfrac{\mathfrak{D}^{n}O\left(p,x\right)}{\partial x^{n}}U\left(Z,x\right)\Rightarrow\left.\dfrac{\partial^{n}O^{\left(Z\right)}}{\partial x^{n}}\right|_{Z=x}=\dfrac{\mathfrak{D}^{n}O\left(p,x\right)}{\partial x^{n}}
\end{equation}
and the limit $Z\rightarrow x$ is taken immediately with $U\left(x,x\right)=1$,
proving the Moyal product \eqref{eq:MOYAL-pure-gauge} in the pure
gauge case. The pure gauge case is in particular usefull to describe
1D situations in the quasi-static limit. In that case, only the gauge
potentials $A_{x}$ and $A_{0}$ might survive, and the electric field
is generated by the covariant derivative of $A_{0}$: $\mathfrak{D}_{x}A_{0}=\partial_{x}A_{0}-\mathbf{i}\left[A_{x},A_{0}\right]$
corresponds to the definition of the quasi-static electric field.

In the case of Abelian field, one simply assumes that all gauge fields
commute with the operators $O_{1,2}$, and that these later commute
among themselves. In addition, the covariant derivatives $\mathfrak{D}_{x}$
collapse to the usual ones $\partial_{x}$. So one has, at leading
order
\begin{multline}
O_{1}\left(p,x\right)\star O_{2}\left(p,x\right)=O_{1}\left(p,x\right)O_{2}\left(p,x\right)+\dfrac{\mathbf{i}}{2}\left(\dfrac{\partial O_{1}}{\partial x^{\alpha}}\dfrac{\partial O_{2}}{\partial p_{\alpha}}-\dfrac{\partial O_{1}}{\partial p_{\alpha}}\dfrac{\partial O_{2}}{\partial x^{\alpha}}-F_{\alpha\beta}\left(x\right)\dfrac{\partial O_{1}}{\partial p_{\alpha}}\dfrac{\partial O_{2}}{\partial p_{\beta}}\right)+\\
+\dfrac{1}{12}\dfrac{\partial F_{\gamma\beta}}{\partial x^{\alpha}}\left(\dfrac{\partial O_{1}}{\partial p_{\gamma}}\dfrac{\partial^{2}O_{2}}{\partial p_{\alpha}\partial p_{\beta}}+\dfrac{\partial^{2}O_{1}}{\partial p_{\alpha}\partial p_{\beta}}\dfrac{\partial O_{2}}{\partial p_{\gamma}}\right)+\mathcal{O}\left(\partial^{4}\right)\label{eq:MOYAL-EXPANSION-ABELIAN}
\end{multline}
as first obtained in \citep{Muller1999a,Lein2010}. Note nevertheless
that, in some cases, the terms are classified in different ways. For
instance, the term $\partial F\cdot\partial_{p}^{2}O_{1}\cdot\partial_{p}O_{2}$
is sometimes thought as a fourth order term in gradient, and thus
comes in regards of terms like $\partial_{x}^{2}O_{1}\cdot\partial_{p}^{2}O_{2}$,
e.g. in \citep{Karasev2003}. We choose here to expand in powers of
the gradients of the operators $O_{1,2}$, rendering the second line
of \eqref{eq:MOYAL-EXPANSION-ABELIAN} a third order term.

We no more elaborate on the limit of Abelian gauge fields, since it
has been investigated in many situations, see e.g. \citep{Luttinger1951,Stratonovich1956,Kelly1964,Kubo1964,Langreth1966,Bialynicki-Birula1977,Serimaa1986,Javanainen1987,Bialynicki-Birula1991,Altshuler1992,Best1993,Levanda1994,Zachos1999,Levanda2001,Swiecicki2013}
and references therein.

\section{Gauge-covariant gradient expansion in the phase-space\label{sec:Gauge-covariant-gradient-expansion-phase-space}}

In all the previous sections, we discussed the situation when the
gauge fields were position dependent. With the upraising of topological
matter, it becomes of utmost importance to understand the possible
non-trivial topologies of the band structure, see e.g. \citep{Hasan2010,Xiao2010,Nagaosa2010,Charge2016}
for reviews. Such non-trivial geometry comes from some gauge structure
in the momentum space, as introduced quickly in section \ref{subsec:Emergence-of-Gauge-structures}.
Despite the possible difficulties to deal with periodic systems in
the Brillouin zone, the topology of the band structure might naively
be thought as a momentum space topology in the phase space. In fact,
naive interpretation along these lines seems to correctly describe
the phenomenology of the already known materials, see e.g. \citep{Xiao2010,Nagaosa2010,Charge2016}
and references therein. In this section, we discuss how to deal with
such structure, when gauge field appears in the form of 
\begin{equation}
F^{\alpha\beta}\left(p\right)=\partial^{\alpha}A^{\beta}\left(p\right)-\partial^{\beta}A^{\alpha}\left(p\right)-\mathbf{i}\left[A^{\alpha}\left(p\right),A^{\beta}\left(p\right)\right]
\end{equation}
instead of \eqref{eq:F-DEFINITION}. We will note all the $p$-derivative
with upper indices as in $\partial^{\alpha}f\equiv\partial f/\partial p_{\alpha}$,
to distinguish easilly $F_{\alpha\beta}\left(x\right)$ from $F^{\alpha\beta}\left(p\right)$. 

\subsection{Momentum-like gauge-covariant gradient expansion\label{subsec:Momentum-like-gauge-covariant-gradient-expansion}}

In this section, we discuss the properties of the gauge-covariant
Moyal expansion in the reciprocal space. Namely, we define another
Wigner transform being covariant in the momentum space instead of
the position space as \eqref{eq:gauge-covariant-Wigner-transform}
was. Since the associated Moyal expansion can be obtained from \eqref{eq:STAR-PRODUCT}
with minor replacements (essentially $x\leftrightarrow p$), we will
be quick with the presentation.

As discussed in section \ref{sec:Gauge-covariant-Moyal-star-product},
the gauge-covariant Wigner transform can be constructed heuristically
from the definition \eqref{eq:Wigner-formal-dx} where we substitute
the flat derivative by covariant ones $\partial_{x}\rightarrow D_{x}=\partial_{x}-\mathbf{i}A_{x}\left(x\right)$,
ultimately giving the parallel transport operators $U\left(x_{1},x_{2}\right)$
in the definition of a covariant Wigner transform \eqref{eq:gauge-covariant-Wigner-transform}.
For the sake of pedagogy, we slightly change the notation here, and
introduce the notation 
\begin{equation}
O^{x}\left(p,x\right)=\int d\mathfrak{x}\left[e^{-\mathbf{i}p\cdot\mathfrak{x}}\left[e^{\mathfrak{x}\cdot D_{x_{1}}/2}O\left(x_{1},x_{2}\right)e^{-\mathfrak{x}\cdot D_{x_{2}}^{\dagger}/2}\right]_{x_{1}=x_{2}=x}\right]\label{eq:Wigner-Ox}
\end{equation}
for the covariant Wigner transform of the operator $O\left(x_{1},x_{2}\right)$
that we discussed all along the previous sections. 

When dealing with band structure calculations, one might prefer to
start from the operator $O\left(p_{1},p_{2}\right)$ in the momentum
representation. Since the momentum representation $O\left(p_{1},p_{2}\right)$
is obtained from the space representation $O\left(x_{1},x_{2}\right)$
via Fourier transformation, one can define as well the Wigner transform
in the alternative form, see \eqref{eq:Wigner},

\begin{multline}
O\left(p,x\right)=\int\dfrac{d\mathfrak{p}}{2\pi}\left[e^{\mathbf{i}\mathfrak{p}\cdot x}\left[e^{\mathfrak{p}\cdot\partial_{p_{1}}/2}O\left(p_{1},p_{2}\right)e^{-\mathfrak{p}\cdot\partial_{p_{2}}^{\dagger}/2}\right]_{p_{1}=p_{2}=p}\right]\\
\Rightarrow O^{p}\left(p,x\right)=\int\dfrac{d\mathfrak{p}}{2\pi}\left[e^{\mathbf{i}\mathfrak{p}\cdot x}\left[e^{\mathfrak{p}\cdot D_{p_{1}}/2}O\left(p_{1},p_{2}\right)e^{-\mathfrak{p}\cdot D_{p_{2}}^{\dagger}/2}\right]_{p_{1}=p_{2}=p}\right]\label{eq:Wigner-Op}
\end{multline}
with $D_{p}=\partial_{p}-\mathbf{i}A\left(p\right)$. We note the
gauge potential in the momentum space using the same symbols as the
gauge potential $A\left(x\right)$ in the real space, their variables
make explicit which one we are dealing with. Despite the two representations
of the flat-space Wigner transforms in \eqref{eq:Wigner}, starting
either from $O\left(x_{1},x_{2}\right)$ or $O\left(p_{1},p_{2}\right)$
and taking the partial Fourier transform with respect to $\mathfrak{x}=x_{1}-x_{2}$
or $\mathfrak{p}=p_{1}-p_{2}$, are strictly equivalent, their covariant
extensions $O^{x}$ and $O^{p}$ are not related by any Fourier transform.
Clearly, $O^{x}$ in \eqref{eq:Wigner-Ox} is well adapted to treat
problem with gauge fields in the position space. By extension, it
becomes clear that $O^{p}$ in \eqref{eq:Wigner-Op} would be of interest
for problems with non-trivial geometry in the momentum space. So in
this section we discuss briefly the phase-space gradient extension
using \eqref{eq:Wigner-Op} instead of \eqref{eq:Wigner-Ox} as the
gauge-covariant Wigner transform.

Nevertheless, discussion can become quite short when realizing that
the property \eqref{eq:U-exp-DEF} works in the same way in either
the $x$ or $p$ spaces. So whenever we had
\begin{equation}
O^{x}\left(p,x\right)=\int d\mathfrak{x}\left[e^{-\mathbf{i}p\cdot\mathfrak{x}}U\left(x,x+\dfrac{\mathfrak{x}}{2}\right)O\left(x+\dfrac{\mathfrak{x}}{2},x-\dfrac{\mathfrak{x}}{2}\right)U\left(x-\dfrac{\mathfrak{x}}{2},x\right)\right]
\end{equation}
in section \ref{sec:Gauge-covariant-Moyal-star-product}, we will
now have 
\begin{equation}
O^{p}\left(p,x\right)=\int\dfrac{d\mathfrak{p}}{2\pi\hbar}\left[e^{\mathbf{i}\mathfrak{p}\cdot x}U\left(p,p+\dfrac{\mathfrak{p}}{2}\right)O\left(p+\dfrac{\mathfrak{p}}{2},p-\dfrac{\mathfrak{p}}{2}\right)U\left(p-\dfrac{\mathfrak{p}}{2},p\right)\right]
\end{equation}
in the present section. In particular, one can write symbolically
\begin{equation}
O^{x}\left(p,x\right)=U\left(x,x+\dfrac{\mathbf{i}}{2}\dfrac{\partial}{\partial p}\right)\int d\mathfrak{x}\left[e^{-\mathbf{i}p\cdot\mathfrak{x}}O\left(x+\dfrac{\mathfrak{x}}{2},x-\dfrac{\mathfrak{x}}{2}\right)\right]U\left(x-\dfrac{\mathbf{i}}{2}\dfrac{\partial^{\dagger}}{\partial p},x\right)
\end{equation}
using the usual Bopp substitution \citep{Polkovnikov2009}, see also
\eqref{eq:pseudo-Bopp-substitution}. We recognise the usual flat-space
Wigner transform inside the formal $U$'s operators. We write thus
in symbolic form
\begin{equation}
O^{x}\left(p,x\right)=U\left(x,x+\dfrac{\mathbf{i}}{2}\dfrac{\partial}{\partial p}\right)O\left(p,x\right)U\left(x-\dfrac{\mathbf{i}}{2}\dfrac{\partial^{\dagger}}{\partial p},x\right)
\end{equation}
and, accordingly, one has
\begin{equation}
O^{p}\left(p,x\right)=U\left(p,p-\dfrac{\mathbf{i}}{2}\dfrac{\partial}{\partial x}\right)O\left(p,x\right)U\left(p+\dfrac{\mathbf{i}}{2}\dfrac{\partial^{\dagger}}{\partial x},p\right)
\end{equation}
for the covariant Wigner transform with momentum non-trivial geometric
structures.

Next, whenever we wanted to construct the covariant Moyal product,
we introduced the mapping \eqref{eq:mapping-flat-covariant} allowing
to transform the $x$-derivatives to some covariant ones. This was
insured by the definition of the $O^{\left(Z\right)}\left(p,x\right)$
Wigner transform \eqref{eq:OZ} which formally looks like the Wigner
transform in the flat space $O\left(p,x\right)$ (in the notation
of this section, $O\left(p,x\right)$ is the non-covariant Wigner
transform \eqref{eq:Wigner}). So the mapping \eqref{eq:mapping-flat-covariant}
can be formally represented as the transformation from the flat space
to the covariant representations, and in the notations of this section,
one would have 
\begin{equation}
O_{1}^{x}\left(p,x\right)\star O_{2}^{x}\left(p,x\right)=U\left(x,x+\dfrac{\mathbf{i}}{2}\dfrac{\partial}{\partial p}\right)\left[O_{1}\left(p,x\right)\star_{0}O_{2}\left(p,x\right)\right]U\left(x-\dfrac{\mathbf{i}}{2}\dfrac{\partial^{\dagger}}{\partial p},x\right)\label{eq:mapping-x}
\end{equation}
instead of the transformation from $O^{\left(Z\right)}$ to $O^{x}$.
Said differently, the mappings \eqref{eq:mapping-flat-covariant}
and \eqref{eq:mapping-x} are equivalent. Finally, the reciprocal
Moyal product would be defined as 
\begin{equation}
O_{1}^{p}\left(p,x\right)\star O_{2}^{p}\left(p,x\right)=U\left(p,p-\dfrac{\mathbf{i}}{2}\dfrac{\partial}{\partial x}\right)\left[O_{1}\left(p,x\right)\star_{0}O_{2}\left(p,x\right)\right]U\left(p+\dfrac{\mathbf{i}}{2}\dfrac{\partial^{\dagger}}{\partial x},p\right)\label{eq:mapping-p}
\end{equation}
allowing to use all the tricks developed in section \ref{subsec:Substitutions-for-the-derivatives}
with only a few adaptations. For instance, instead of \ref{eq:substitution-dOZ-DO-expansion},
we will have now 
\begin{multline}
U\left(p,p-\dfrac{\mathbf{i}}{2}\dfrac{\partial}{\partial x}\right)\dfrac{\partial O}{\partial p_{\alpha}}U\left(p+\dfrac{\mathbf{i}}{2}\dfrac{\partial^{\dagger}}{\partial x},p\right)=\\
=\dfrac{\mathfrak{D}O\left(p,x\right)}{\partial p_{\alpha}}+\left[\sum_{n=0}^{\infty}\dfrac{\mathbf{i}^{n}\left(z\cdot\mathfrak{D}_{x}\right)^{n}}{2^{n+1}\left(n+2\right)!}F^{\alpha\beta}\left(p\right)z_{\beta},O\left(p,x\right)\right]_{z=\partial_{x}-\partial_{x}^{\dagger}}\label{eq:susbtitution-dOZ-dp}
\end{multline}
for the replacement of the first order $p$-derivative, with 
\begin{equation}
\dfrac{\mathfrak{D}O}{\partial p_{\alpha}}=\dfrac{\partial O}{\partial p_{\alpha}}-\mathbf{i}\left[A^{\alpha}\left(p\right),O\left(p,x\right)\right]\;;\;F^{\alpha\beta}\left(p\right)=\dfrac{\partial A^{\beta}}{\partial p_{\alpha}}-\dfrac{\partial A^{\alpha}}{\partial p_{\beta}}-\mathbf{i}\left[A^{\alpha}\left(p\right),A^{\beta}\left(p\right)\right]
\end{equation}
so we essentially replace the covariant derivative by contravariant
ones, and raise up the indices, keeping in mind the usual symplectic
structure of the phase space.

We do not give the gradient expansion of the Moyal product, since
it is mutadis mutandis the one given in \eqref{eq:STAR-PRODUCT}:
instead of replacing the $x$-derivatives in the $\star_{0}$-product
and keeping the $p$-derivatives in section \ref{sec:Gauge-covariant-Moyal-star-product},
we now replace the $p$-derivatives in the $\star_{0}$-product, and
let the $x-$derivative untouched. For the first derivative the substitution
reads \eqref{eq:susbtitution-dOZ-dp}.

We conclude this section with a few remarks about the gauge structure
of the momentum space. Firstly, its relation to the periodic band
structure is not yet established, a task kept for later studies. A
few first steps were done in \citep{Zubkov2016}, and it seems at
first to be not much problematic, at least from a physicist point
of view. Its rigorous mathematical establishment might be more complex,
though. Secondly, we did not discuss the origin of the gauge potential
$A\left(p\right)$ and its covariant generalization to the gauge field
$F^{\alpha\beta}\left(p\right)$. As in the case of the $x$-space
non-trivial geometry, the operators appearing in the Moyal product
\eqref{eq:STAR-PRODUCT} would be the bare ones. So, an understanding
of the non-trivial geometry in the momentum space must be done at
an other, perhaps more microscopic, level. Reader interested in the
topology of the band structure might consult \citep{Shapere1989a,Bohm2003,Chruscinski2004}
and references therein, as well as \citep{Hasan2010,Xiao2010,Nagaosa2010,Charge2016}.
Recent studies \citep{Stephanov2012,Son2012,Son2013,Son2013a,Stephanov2016}
discovered how the Berry connection associated to the band structure
of Weyl semi-metal appears at the quasi-classical level using non-covariant
gradient expansions of a topological action, see also \citep{Duval2006a,Duval2006b,Stone2013}
and references therein for a bit more mathematically oriented approach.
Other approaches use the deformed wave-packet dynamics and its adiabatic
corrections to get such phenomenologies, see \citep{Xiao2010} for
a review.

\subsection{Phase-space gradient expansion in a covariant manner\label{subsec:Phase-space-gradient-expansion}}

Section \ref{sec:Gauge-covariant-Moyal-star-product} was devoted
to the construction of a covariant method to generalise the Moyal
product to non-Abelian gauge structure in the position space, and
section \ref{subsec:Momentum-like-gauge-covariant-gradient-expansion}
showed how it is possible to adapt such formalism to the situation
when the system present non-trivial geometry in the momentum space.
None of the gradient expansions developed so far present a complete
symmetry of position and momentum, since the gauge structure was non-trivial
only in either the position or the momentum space, not in the entire
phase space. I would like to bring a few arguments in favor of a generalisation
of the two above approaches towards a phase-space gradient expansion,
in a fully covariant manner at all stages of the calculation.

To explore such a possibility, we start from the Dyson's equation
\begin{equation}
\int dy\left[G^{-1}\left(x_{1},y\right)G\left(y,x_{2}\right)\right]=\delta\left(x_{1}-x_{2}\right)
\end{equation}
which transforms under the transformation \eqref{eq:Wigner-Ox} and
the mapping \eqref{eq:mapping-x} as 
\begin{multline}
U_{x}\left[G^{-1}\left(p,x\right)\star_{0}G\left(p,x\right)\right]\bar{U}_{x}=1\\
\text{with}\;\;U_{x}=U\left(x,x+\dfrac{\mathbf{i}}{2}\dfrac{\partial}{\partial p}\right)\;\;;\;\;\bar{U}_{x}=U\left(x-\dfrac{\mathbf{i}}{2}\dfrac{\partial^{\dagger}}{\partial p},x\right)
\end{multline}
with $G\left(p,x\right)$ the Wigner transform \eqref{eq:Wigner}
of the Green's function. We have similarly 
\begin{multline}
U_{p}\left[G^{-1}\left(p,x\right)\star_{0}G\left(p,x\right)\right]\bar{U}_{p}=1\\
\text{with}\;\;U_{p}=U\left(p,p-\dfrac{\mathbf{i}}{2}\dfrac{\partial}{\partial x}\right)\;\;;\;\;\bar{U}_{p}=U\left(p+\dfrac{\mathbf{i}}{2}\dfrac{\partial^{\dagger}}{\partial x},p\right)
\end{multline}
using the transformation \eqref{eq:Wigner-Op} and the mapping \eqref{eq:mapping-p}
being covariant in the momentum space. Because we choose to transform
the Dyson's equation, the right-hand-side is always the identity matrix,
and so nothing forbids to apply the two mappings \eqref{eq:mapping-x}
and \eqref{eq:mapping-p} succesively, i.e. to define
\begin{equation}
U_{p}U_{x}\left[G^{-1}\left(p,x\right)\star_{0}G\left(p,x\right)\right]\bar{U}_{x}\bar{U}_{p}=1
\end{equation}
as the phase-space representation of the Dyson's equation. At first
order in the gradient, one has 
\begin{equation}
\dfrac{\partial}{\partial x^{\alpha}}\overset{\mathcal{U}_{x}}{\longrightarrow}U_{x}\dfrac{\partial}{\partial x^{\alpha}}\bar{U}_{x}=\dfrac{\mathfrak{D}}{\partial x^{\alpha}}+\dfrac{1}{4}\left\{ F_{\alpha\beta}\left(x\right),\dfrac{\partial}{\partial p_{\beta}}\right\} +\cdots
\end{equation}
\begin{equation}
\dfrac{\partial}{\partial p_{\alpha}}\overset{\mathcal{U}_{p}}{\longrightarrow}U_{p}\dfrac{\partial}{\partial p_{\alpha}}\bar{U}_{p}=\dfrac{\mathfrak{D}}{\partial p_{\alpha}}+\dfrac{1}{4}\left\{ F^{\alpha\beta}\left(p\right),\dfrac{\partial}{\partial x^{\beta}}\right\} +\cdots
\end{equation}
whereas $\partial_{x}$ is an invariant of the $\mathcal{U}_{p}$
mapping and $\partial_{p}$ is invariant under the $\mathcal{U}_{x}$
mapping. So applying only one time $\mathcal{U}_{x,p}$ on the Dyson's
equation does not generate fully gauge covariant expressions in the
full phase-space. To remedy this problem, we apply as many $\mathcal{U}_{x,p}$
operations than one needs. For instance, at first order in gradient,
and second order in fields, one would have 

\begin{align}
\dfrac{\partial}{\partial x^{\alpha}} & \rightarrowtriangle\dfrac{\mathfrak{D}}{\partial x^{\alpha}}+\dfrac{1}{4}\left\{ F_{\alpha\beta}\left(x\right),\dfrac{\mathfrak{D}}{\partial p_{\beta}}\right\} +\dfrac{1}{16}\left\{ F_{\alpha\beta}\left(x\right),\left\{ F^{\beta\gamma}\left(p\right),\dfrac{\mathfrak{D}}{\partial x^{\gamma}}\right\} \right\} +\cdots\nonumber \\
\dfrac{\partial}{\partial p_{\alpha}} & \rightarrowtriangle\dfrac{\mathfrak{D}}{\partial p_{\alpha}}+\dfrac{1}{4}\left\{ F^{\alpha\beta}\left(p\right),\dfrac{\mathfrak{D}}{\partial x^{\beta}}\right\} +\dfrac{1}{16}\left\{ F^{\alpha\beta}\left(p\right),\left\{ F_{\beta\gamma}\left(x\right),\dfrac{\mathfrak{D}}{\partial p_{\gamma}}\right\} \right\} +\cdots\label{eq:mapping-covariant-phase-space}
\end{align}
and we note this mapping with the special arrow $\rightarrowtriangle$
for commodity. We confess the mapping is clearly not well defined
that way, but we give heuristic arguments for its construction. Clearly,
the mapping \eqref{eq:mapping-covariant-phase-space} requires an
extra scale than the usual gradient expansion to be truncated at the
desired order. Here, we again used heuristic arguments, truncating
it at  first order in gradient and second order in gauge fields in
an arbitrary manner. Nevertheless, the mapping \eqref{eq:mapping-covariant-phase-space}
is clearly covariant in the entire phase-space, as required.

Using the heuristic mapping \eqref{eq:mapping-covariant-phase-space},
one obtains for the gradient expansion of the Dyson equation, which
we note with a big star
\begin{multline}
G^{-1}\left(p,x\right)\bigstar G\left(p,x\right)=G^{-1}\left(p,x\right)G\left(p,x\right)+\dfrac{\mathbf{i}}{2}\left(\dfrac{\mathfrak{D}G^{-1}}{\partial x^{\alpha}}\dfrac{\mathfrak{D}G}{\partial p_{\alpha}}-\dfrac{\mathfrak{D}G^{-1}}{\partial p_{\alpha}}\dfrac{\mathfrak{D}G}{\partial x^{\alpha}}\right)-\\
-\dfrac{\mathbf{i}}{8}\left(F_{\alpha\beta}\left(x\right)\dfrac{\mathfrak{D}G^{-1}}{\partial p_{\alpha}}\dfrac{\mathfrak{D}G}{\partial p_{\beta}}+2\dfrac{\mathfrak{D}G^{-1}}{\partial p_{\alpha}}F_{\alpha\beta}\left(x\right)\dfrac{\mathfrak{D}G}{\partial p_{\beta}}+\dfrac{\mathfrak{D}G^{-1}}{\partial p_{\alpha}}\dfrac{\mathfrak{D}G}{\partial p_{\beta}}F_{\alpha\beta}\left(x\right)-\left(\begin{array}{c}
\mathfrak{D}_{p}\rightarrow\mathfrak{D}_{x}\\
F_{\alpha\beta}\left(x\right)\rightarrow F^{\alpha\beta}\left(p\right)
\end{array}\right)\right)-\\
-\dfrac{\mathbf{i}}{32}\left(\left\{ F_{\alpha\beta},\dfrac{\mathfrak{D}G^{-1}}{\partial p_{\alpha}}\right\} \left\{ F^{\beta\gamma},\dfrac{\mathfrak{D}G}{\partial x^{\gamma}}\right\} +\left\{ F^{\beta\gamma},\left\{ F_{\alpha\beta},\dfrac{\mathfrak{D}G^{-1}}{\partial p_{\alpha}}\right\} \right\} \dfrac{\mathfrak{D}G}{\partial x^{\gamma}}+\dfrac{\mathfrak{D}G^{-1}}{\partial p_{\alpha}}\left\{ F_{\alpha\beta},\left\{ F^{\beta\gamma},\dfrac{\mathfrak{D}G}{\partial x^{\gamma}}\right\} \right\} \right)+\\
+\dfrac{\mathbf{i}}{32}\left(\left\{ F^{\alpha\beta},\dfrac{\mathfrak{D}G^{-1}}{\partial x^{\alpha}}\right\} \left\{ F_{\beta\gamma},\dfrac{\mathfrak{D}G}{\partial p_{\gamma}}\right\} +\left\{ F_{\beta\gamma},\left\{ F^{\alpha\beta},\dfrac{\mathfrak{D}G^{-1}}{\partial x^{\alpha}}\right\} \right\} \dfrac{\mathfrak{D}G}{\partial p_{\gamma}}+\dfrac{\mathfrak{D}G^{-1}}{\partial x^{\alpha}}\left\{ F^{\alpha\beta},\left\{ F_{\beta\gamma},\dfrac{\mathfrak{D}G}{\partial p_{\gamma}}\right\} \right\} \right)\\
+\dfrac{\mathbf{i}}{32}\left(\begin{array}{c}
\mathfrak{D}_{p}\leftrightarrow\mathfrak{D}_{x}\\
F_{\alpha\beta}\left(x\right)\rightarrow F^{\alpha\beta}\left(p\right)\\
F^{\beta\gamma}\left(p\right)\rightarrow F_{\beta\gamma}\left(x\right)
\end{array}\right)+\cdots=1\label{eq:BIGSTAR-PRODUCT}
\end{multline}
where the indices of the derivatives are conserved under the substitution
$\mathfrak{D}_{p}\leftrightarrow\mathfrak{D}_{x}$. 

We have been able to justify the multiple applications of the transformations
$U_{x,p}$ on the derivatives because the right-hand-side of the Dyson's
equation is the unit matrix, which is an invariant of the $U_{x,p}$'s.
Nevertheless, one sees easilly that \eqref{eq:BIGSTAR-PRODUCT} reduces
to the covariant $\star$-product established in section \ref{sec:Gauge-covariant-Moyal-star-product}
when $F^{\alpha\beta}\left(p\right)\rightarrow0$, and incidentaly
it reduces to the $\star$-product established in section \ref{subsec:Momentum-like-gauge-covariant-gradient-expansion}
in the limit $F_{\alpha\beta}\left(x\right)\rightarrow0$. We thus
promote -- without more justification -- the $\bigstar$-product
\eqref{eq:BIGSTAR-PRODUCT} as an elligible Moyal product in the phase-space.
We now study a few immediate consequences of this Moyal algebra, and
see that it completes some previoulsy calculated star products appearing
in the study of effective theories.

From \eqref{eq:BIGSTAR-PRODUCT}, replacing $G^{-1}$ and/or $G$
with some variables $x^{\alpha}$ and/or $p_{\beta}$, one gets 
\begin{align}
\left[x^{\alpha},p_{\beta}\right]_{\bigstar} & =x^{\alpha}\bigstar p_{\beta}-p_{\beta}\bigstar x^{\alpha}=\mathbf{i}\delta_{\beta}^{\alpha}+\dfrac{3\mathbf{i}}{8}\left\{ F^{\alpha\gamma}\left(p\right),F_{\gamma\beta}\left(x\right)\right\} \nonumber \\
\left[x^{\alpha},x^{\beta}\right]_{\bigstar} & =x^{\alpha}\bigstar x^{\beta}-x^{\beta}\bigstar x^{\alpha}=\mathbf{i}F^{\alpha\beta}\left(p\right)\nonumber \\
\left[p_{\alpha},p_{\beta}\right]_{\bigstar} & =p_{\alpha}\bigstar p_{\beta}-p_{\beta}\bigstar p_{\alpha}=-\mathbf{i}F_{\alpha\beta}\left(x\right)\label{eq:big-star-commutation}
\end{align}
for the non-commutative algebra of the phase-space. Previous works
essentially discussed either the momentum space covariant construction
(when $F_{\alpha\beta}=0$) being non trivial when monopoles are present
in the band structures \citep{Sundaram1999,Duval2001,Berard2004}
(see also \citep{Xiao2010,Volovik2003} for reviews) or the position
space covariant construction (when $F^{\alpha\beta}=0$) being non
trivial under magnetic field -- this is the famous Landau problem
important for the quantum Hall effect \citep{Bellissard1994}. The
price to pay to associate both effects is the higher order term $\left\{ F^{\alpha\gamma}\left(p\right),F_{\gamma\beta}\left(x\right)\right\} $
appearing in the dynamics. This term has been identified in the past
in the Abelian limit \citep{Xiao2005,Duval2006b,Duval2006a,Bliokh2006},
and it is responsible for some characteristic phenomenologies of the
Weyl semimetals, like the anomalous Hall effect (see \citep{Nagaosa2010}
for a review on related phenomena), the chiral anomaly \citep{Stephanov2012,Son2012,Son2013}
and the negative magneto-resistance \citep{Son2013a}.

One can as well calculate the equations of motion
\begin{equation}
\mathbf{i}\dfrac{\partial f}{\partial t}=\left[f\left(p,x\right),H\right]_{\bigstar}=f\bigstar H-H\bigstar f
\end{equation}
for $x^{\alpha}$ and $p_{\alpha}$. One gets
\begin{multline}
\dfrac{\partial x^{\alpha}}{\partial t}=\dfrac{\mathfrak{D}H}{\partial p_{\alpha}}+\dfrac{1}{2}\left\{ F^{\alpha\beta}\left(p\right),\dfrac{\mathfrak{D}H}{\partial x^{\beta}}\right\} \\
+\dfrac{1}{16}\left(2\left\{ F^{\alpha\beta}\left(p\right),\left\{ F_{\beta\gamma}\left(x\right),\dfrac{\mathfrak{D}H}{\partial p_{\gamma}}\right\} \right\} +\left\{ \left\{ F^{\alpha\beta}\left(p\right),F_{\beta\gamma}\left(x\right)\right\} ,\dfrac{\mathfrak{D}H}{\partial p_{\gamma}}\right\} \right)\label{eq:x-dot}
\end{multline}
and
\begin{multline}
\dfrac{\partial p_{\alpha}}{\partial t}=-\dfrac{\mathfrak{D}H}{\partial x^{\alpha}}-\dfrac{1}{2}\left\{ F_{\alpha\beta}\left(x\right),\dfrac{\mathfrak{D}H}{\partial p_{\beta}}\right\} \\
-\dfrac{1}{16}\left(2\left\{ F_{\alpha\beta}\left(x\right),\left\{ F^{\beta\gamma}\left(p\right),\dfrac{\mathfrak{D}H}{\partial x^{\gamma}}\right\} \right\} +\left\{ \left\{ F_{\alpha\beta}\left(x\right),F^{\beta\gamma}\left(p\right)\right\} ,\dfrac{\mathfrak{D}H}{\partial x^{\gamma}}\right\} \right)\label{eq:p-dot}
\end{multline}
which reduces to the usual Hamilton relations whenever $A_{\alpha}\left(x\right)=A^{\alpha}\left(p\right)=0$.
In the quasi-static limit, one can replace $H\rightarrow\mathcal{E}-A_{0}\left(p\right)-A_{0}\left(x\right)$
with some time-sector gauge potential $A_{0}$ and energymatrix $\mathcal{E}$.
In that case, the covariant derivatives $\mathfrak{D}_{x}H$ and $\mathfrak{D}_{p}H$
in \eqref{eq:x-dot} and \eqref{eq:p-dot} generate some quasi-static
electric-like fields $F_{0\alpha}\left(x\right)$ and/or $F^{\alpha0}\left(p\right)$. 

When the second lines of \eqref{eq:x-dot} and \eqref{eq:p-dot} are
neglected and the gauge fields are supposed Abelian, these equations
of motion have been thoroughly studied \citep{Gosselin2006,Gosselin2008,Shindou2008,Gosselin2009,Wong2011,Wickles2013},
and often noted as $F^{\alpha\beta}=\varepsilon^{\alpha\beta\gamma}\Omega_{\gamma}$
in terms of the Berry curvature $\Omega_{\gamma}$, whereas $F_{\alpha\beta}=\varepsilon_{\alpha\beta\gamma}B^{\gamma}$
with the magnetic field $B^{\gamma}$. Nevertheless, some studies
found an extra contribution \citep{Shindou2005,Wong2011,Wickles2013},
which could be written as $F\left(p,x\right)$ in the notation of
this paper. I have no idea how such contributions might come out from
the method presented in this paper. However, it is clear that the
method used in \citep{Shindou2005,Wong2011,Wickles2013} is intrinsically
non-covariant, hence its elaboration is by far more intricate than
the above methodology. For instance, many authors write the Berry
connection as, e.g. $A_{p}=U\partial_{p}U^{\dagger}$ with $U$ a
unitary matrix. This is absurd providing such a \textit{connection}
has no intrinsic curvature... A rigorous mathematical construction
has been nevertheless given in \citep{Gosselin2009} and also exhibit
such $x-p$ term. Further elaborations will be presented in later
studies.

Here I gave a(n almost) clear justification of the $x$ and $p$ gauge
structure, and explained how one can elaborate a covariant semiclassical
method, starting from a classical expression which is later dressed
by the gradient expansion $\bigstar$ \eqref{eq:BIGSTAR-PRODUCT}.

\section{Conclusion}

Though it might appear as being not rigorously grounded, the present
study enlight a few characteristics of the non-commutative geometry
in the phase space. In particular, a general method has been developed
to generate a gauge-covariant gradient expansion order by order in
section \ref{sec:Gauge-covariant-Moyal-star-product}. It is based
on a formal mapping between the flat gauge Moyal product and the gauge
covariant one introduced in section \ref{subsec:Two-different-Wigner-transforms}.
In particular, the mapping between position derivatives in the two
representations has been established, at least for the first few orders,
in section \ref{subsec:Substitutions-for-the-derivatives}. When dealing
with momentum space gauge structures (see section \ref{subsec:Emergence-of-Gauge-structures}
for a quick introduction), the mapping applies instead on the momentum
derivatives, see section \ref{subsec:Momentum-like-gauge-covariant-gradient-expansion}.
It has been shown that these mappings generate only covariant contributions
to the Moyal star-product, unlike the usual approach intensively used
in the physics literature. 

Such star products automatically dress the classical theory with the
gauge structure, in addition to provide the usual quantum corrections
to the classical statistics in the form of the gradient expansion.
Moreover, one can calculate many usefull properties of statistical
systems using the covariant $\star$-product, like e.g. the linear
response theory, which eventually turn out to be of topological origin,
see section \ref{subsec:Topology-in-the-momentum-space}. It is also
usefull to extract low energy effective theories in a covariant way,
in addition to construct some covariant quantum transport equations
in a quite easy fashion, see section \ref{subsec:Quantum-kinetic-theory}.
These kinetic theories are able to deal with any internal degree of
freedom, like charge, spin, color, ... 

A heuristic generalisation of these star products (covariant either
in the momentum or in the position spaces) has been proposed in section
\ref{subsec:Phase-space-gradient-expansion}, when the gauge structure
is supposed to emerge in both position and momentum spaces. The former
one might come from e.g. internal degree of freedom redundancy, whereas
the gauge structure in the momentum space can emerge from non-trivial
band structure in solid-state systems. Terms coupling both the $x$
and $p$ gauge fields naturally appear in this way. This approach
might be usefull for dealing with the emerging topological condensed
matter systems. Many progresses were done in the past few years, using
non-covariant methods. One can hope that the covariant method proposed
here will be able to unite many (if not all) of the emerging phenomenologies
with the old established ones. This might be a crucial point towards
the complete understanding of the novel states of matter, because
the proposed method present both versatility and easiness of utilisation.

Only part of the phenomenology associated to the generic method presented
in this study has been studied so far (see sections \ref{sec:Moyal-product-in-action}
and \ref{subsec:Topology-in-the-momentum-space} for short reviews),
and many novel effects will be studied in the sequels of this study.
As a matter of fact, I generalised many studies to the non-Abelian
structure. In addition, I constructed a method being covariant at
any level, whereas previous studies made non-covariant constructions
and find bona-fide covariant expansion. Immediate consequences of
the presented star products will be discussed in term of linear-response
theory and covariant construction of effective models in the future.

I hope this study might be of interest for future more rigorous demonstrations
of the gradient expansion in the phase-space, in addition to attract
the attention on a convenient tool to study either high-energy or
consensed matter modern problems.

\bibliography{/home/gozat/Desktop/library}

\end{document}